\pdfoutput=1

\documentclass[aps,showpacs,preprintnumbers,amsmath,amssymb,
eqsecnum, twocolumn, tightenlines
]{revtex4}

\usepackage{graphicx}

\newcommand{\be}{\begin{eqnarray}}
\newcommand{\ee}{\end{eqnarray}}

 \newcommand{\gsim}{\mathrel{\hbox{\rlap{\lower.55ex \hbox {$\sim$}}
                   \kern-.3em \raise.4ex \hbox{$>$}}}}
\newcommand{\lsim}{\mathrel{\hbox{\rlap{\lower.55ex \hbox {$\sim$}}
                   \kern-.3em \raise.4ex \hbox{$<$}}}}

\newcommand{\ba}{\begin{eqnarray}}
\newcommand{\ea}{\end{eqnarray}}

\def\roughly#1{\mathrel{\raise.3ex\hbox{$#1$\kern-.75em%
\lower1ex\hbox{$\sim$}}}}
\def\lsim{\roughly<}
\def\gsim{\roughly>}

\begin{document}


\title{  The  Fate of the Initial State Fluctuations in Heavy Ion Collisions. \\ II
 Glauber Fluctuations and Sounds}
\author {Pilar Staig and Edward Shuryak}
\address { Department of Physics and Astronomy, State University of New York,
Stony Brook, NY 11794}
\date{\today}

\begin{abstract}
Heavy ion collisions at RHIC are  well described by the (nearly
ideal) hydrodynamics for average events. In the present paper we
study initial  state fluctuations appearing on event-by-event
basis, and the propagation of perturbations induced by them. We
found that (i) fluctuations of several lowest harmonics have
comparable magnitudes, (ii) that at least all odd harmonics are
correlated in phase, (iii) thus indicating the local nature of
fluctuations.
 We argue that such local perturbation should be the source of the ``Tiny Bang", a pulse of sound propagating from it.
We identify its two fundamental  scales as (i) the
 ``sound
horizon" (analogous to the absolute ruler in cosmic microwave
background and galaxy distribution)
and (ii) the ``viscous horizon", separating damped and undamped harmonics.
 We then  qualitatively describe how one can determine them from the data,
and thus determine two fundamental parameters of the matter, the (average) {\em speed of sound} and {\em viscosity}.

\end{abstract}
\maketitle

\section{Introduction}

  Starting the introduction, let us note
that  the issues to be discussed in this paper are somewhat
similar in nature to current trends  in cosmology of the last
decade. While the very existence of cosmic microwave background
radiation had dramatically confirmed the existence of the Big Bang
already in the 1960's,  it is its more recent observations which
made cosmology a really quantitative science. Small temperature
fluctuations  on top of the overall Hubble expansion  have been
seen on the sky. Their
 angular size and the magnitude of various harmonics tell us, in surprising detail, about
  sounds propagating the Universe at the plasma neutralization time,   providing precise timing of
the cosmological expansion.

 Experimental data obtained in heavy ion collisions at the Relativistic Heavy Ion Collider  (RHIC)
has found the ``Little Bang", a  hydrodynamical explosion driven
mostly by pressure of the new form of matter, Quark-Gluon Plasma.
Their experimental data for radial and elliptic flows has been
compiled in the so called ``white papers" of all the experiments
in 2004, and compared with predictions of relativistic
hydrodynamics. Very recent results from the Large Hadron Collider
(LHC) on elliptic flow  \cite{ALICE}  also turned out to be in
agreement with hydrodynamical predictions, suggesting that QGP
remains a good liquid even at LHC (see e.g. \cite{viewpoint}).
(For clarity, we mean hydrodynamics complemented by the hadronic
cascade for a correct account of the hadronic stage, see
\cite{hydro,Hirano:2004ta,Nonaka:2006yn}. Models which do not do
that and use rather arbitrary freezeout don't get the energy
dependence
 right, neither for RHIC nor for LHC.)
 Dissipative effects
from the QGP viscosity provide only small corrections at the few percent
level, see
  \cite{Romatschke:2007mq,Dusling:2007gi,Heinz:2008qm}.
 So, by now, we have a good quantitative description of the ``Little Bang".

This paper is however not about it, but about phenomena which we
will call collectively ``The Tiny Bang". In the hydrodynamical
studies mentioned above  one ignored any fluctuations in the
system, using average smooth initial conditions, possessing the
symmetries of the average collisions. (For example, central
collisions are azimuthally symmetric, with vanishing nonzero
Fourier components of the flow. Non-central collisions  were
assumed to produce reflection symmetric $\vec x \rightarrow -\vec
x$ distributions at midrapidity, thus only even harmonics are to
be nonzero.) Yet local perturbation of those should produce extra
excitation of the ``Little Bang",  that should not in general
respect such symmetries.

    Such $perturbations$ of the average explosion can come from at
least two different sources. The one which we will study in this
paper is due to quantum fluctuations in the wave function of the
colliding nuclei, which  creates ``bumpy" distributions of matter,
for any collision, which one can decompose into a smooth average
one plus local perturbations.

Another one, to be studied in subsequent papers of this series,
are created by the energy deposited by  jets propagating through
the medium. It has been recently dramatically shown by ATLAS
collaboration \cite{ATLAS} that even jets with energy above $100
\, GeV$  deposit  large part of their energy, and sometimes all of
it, into the medium. The first ideas were to look at the resulting
perturbations in the form of a  Mach cone \cite{conical}, driven
by  the view that the energy is deposited more or less
homogenously along the jet path.  However more recent developments
of the theory, based on AdS/CFT, have lead to the  view that the
deposited energy grows as the cube of the distance travelled by
the jet, $\Delta E\sim L^3$, with significant deposition near the
endpoint. Thus one may think of the {\em second kind of the ``Tiny
Bangs"}, this time occurring  roughly half time between the
beginning and the end of the ``Little Bang".

\begin{figure}[h]
\begin{center}
\includegraphics*[height=6.cm]{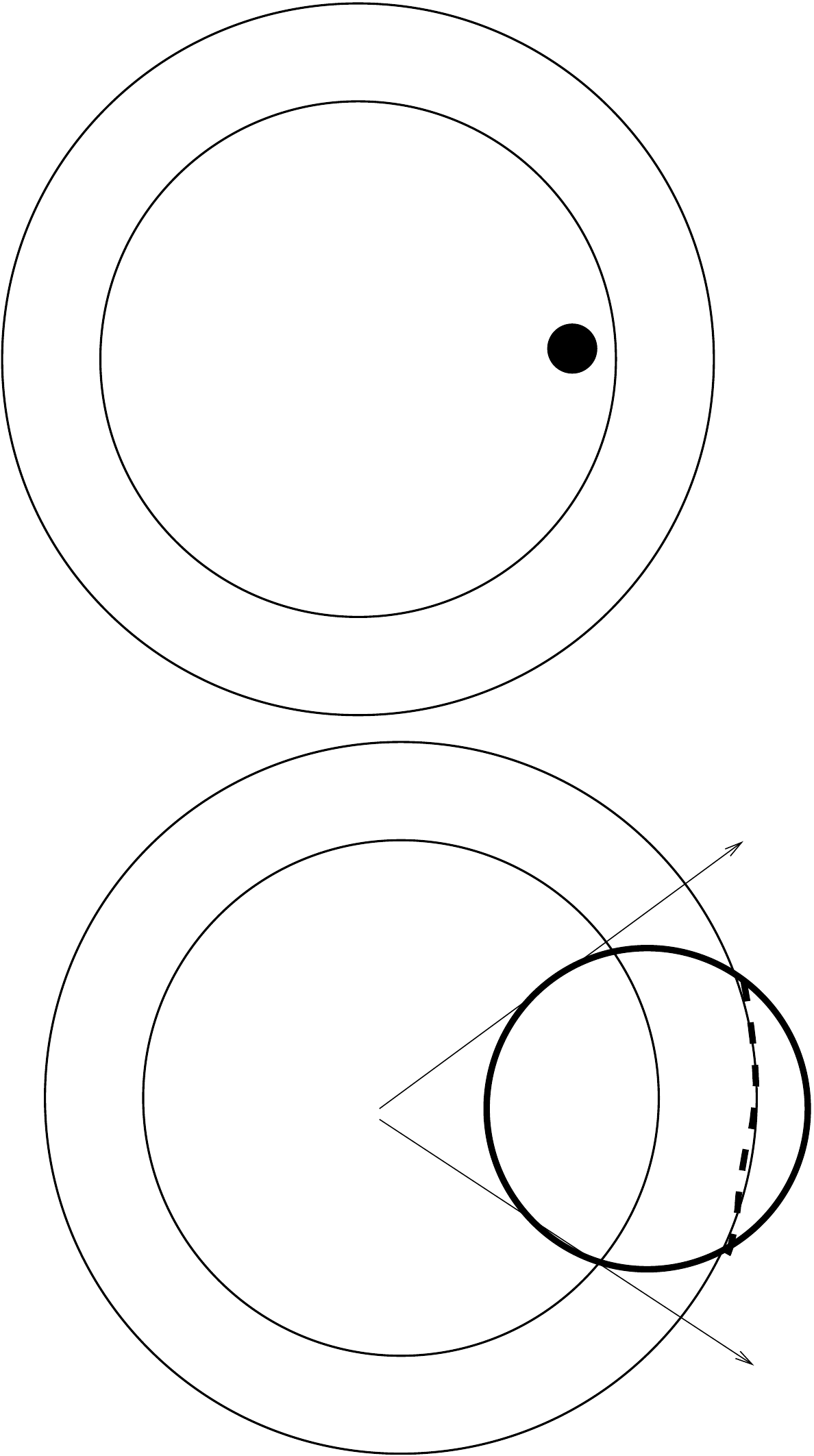}
\caption{A sketch of the transverse plane of the colliding system:
the two concentric circles  are the nuclear radius (inner) and the
final radius of the fireball (outer). The  black spot in the upper
figure is  an extra density at formation time, due to initial
density fluctuations in the collision.
 The small perturbation becomes a circle of a sound wave, which will be stopped at freezeout at thick circle (distorted by radial flow).
The part outside of the fireball does not exist: the corresponding
matter will actually be placed near the edge of the fireball
(thick dashed line). The whole perturbation is enclosed in a
sector between the two thin lines with arrows. }
\label{fig_sketch}
\end{center}
\end{figure}

    The smallness of the perturbation amplitude, with respect to the local
density of ambient matter, would suggest the appearance of
divergent sound  waves, see Fig.\ref{fig_sketch}. Similar to the
circles from a stone thrown into a pond, hydrodynamics tells us
that initial perturbations should become moving waves, with
basically nothing left at the original location at later time.
This is the picture we are going to work on in this paper.

    An alternative idea, of randomly fluctuating shapes of the
produced initial fireballs, has resulted in an approach in which
different angular harmonics of that distribution are treated
separately. The realization that even central $\vec b=0$
collisions may have some fluctuating ellipticity has lead to the
discussion of the elliptic flow event-by-event fluctuations, see
\cite{Mrowczynski:2002bw} and many subsequent works. The so called
``triangular flow" related to the 3-ed harmonic of the flow has
been recently studied by Alver and Roland \cite{Alver:2010gr},
with several groups working in this direction now.

The main difference between our approach and that is that we treat
such fluctuations not as independent noise in different harmonics
but as certain local perturbations, resulting in certain evolving
sound fronts, reaching certain size, shape and diffusivity by the
moment of freezeout. On one hand, one may argue that as soon as
all the perturbations are small and the equations are linear, it
is not important if one expands in harmonics before or after the
solution of hydro equations. And yet we believe that our approach
is not only more intuitive and provides better insights, but it
corresponds to the physical nature of the fluctuations in
question, as higher angular harmonics are in fact correlated with
each other as we show below. Furthermore, we think that our
approach is in more direct connection with the fundamental scales
of the problem, to be detailed below.

Having outlined the main ideas of this paper, let us discuss a bit
more some recent relevant papers. In the last few years RHIC
experiments have focused more on two and three-particle
correlations, which revealed a rather rich phenomenology of
correlations. We are not however able to review those, and we will
refer to data as we develop the theory more.

\begin{figure}[h]
\begin{center}
\includegraphics*[width=7.cm]{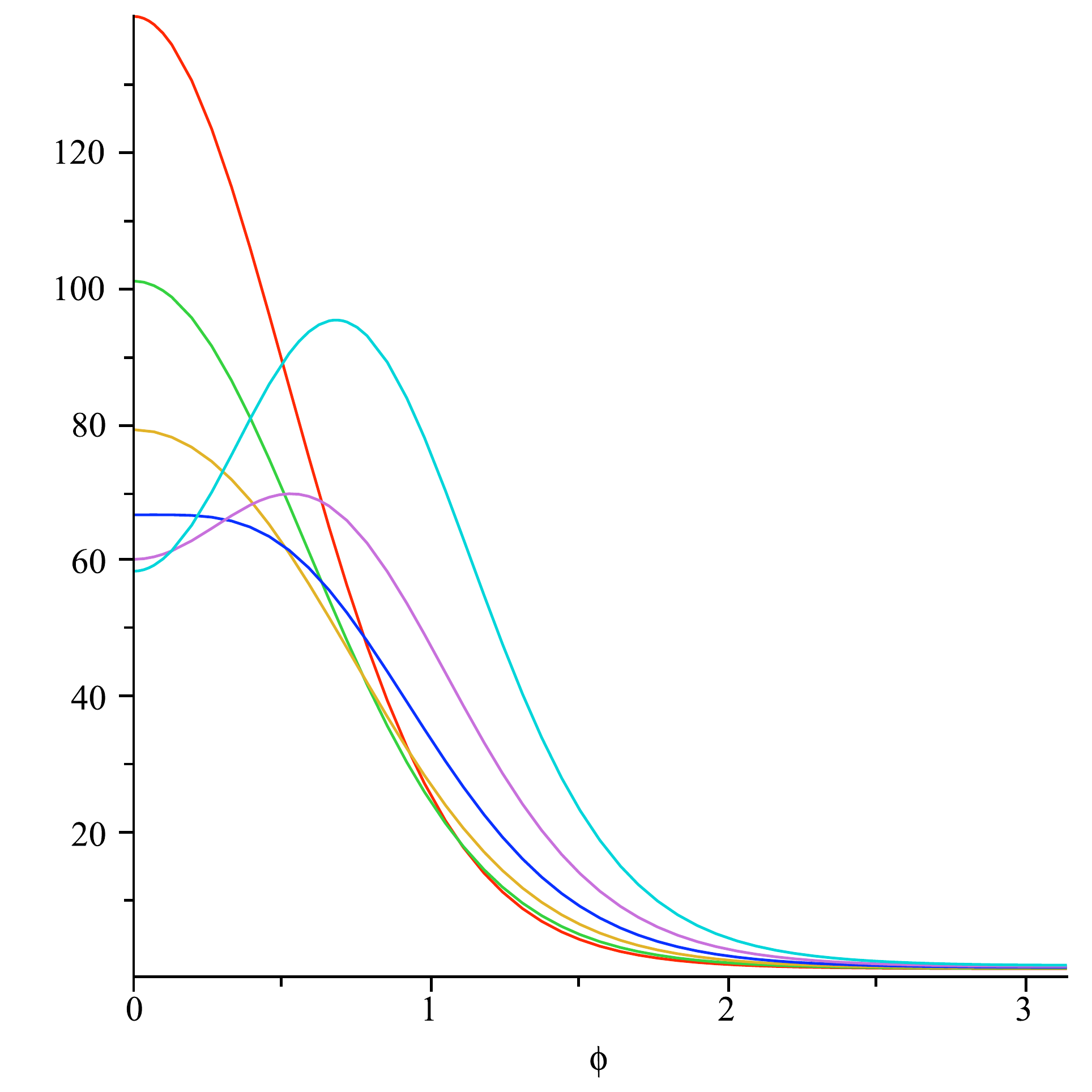}\\
\includegraphics*[width=7.cm]{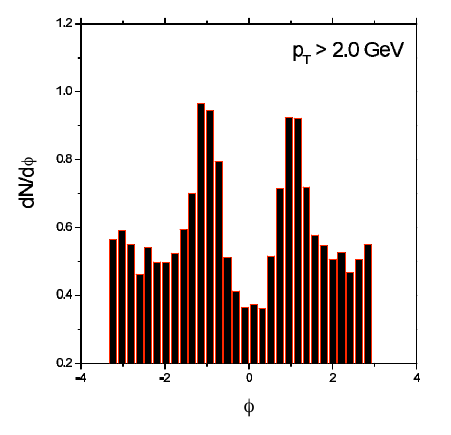}
\caption{Extra particle  distribution in azimuthal angle relative
to position of the initial perturbation. Fig.(a) from
\cite{Shuryak:2009cy} has six curves, from the most narrow to
wider ones, correspond to the radius of the circle 1,2,3,4,5,6 fm,
respectively: the last (blue) with a maximum away from the
original position, corresponds to the sound horizon. The original
spot position is selected to be at the edge of the nuclei and
the observed particle is at $p_t=1\, GeV$.\\
 Fig.(b) from   \cite{Andrade:2009em}, for a particle with $p_t>2\, GeV$
} \label{fig_twopeaks}
\end{center}
\end{figure}

    The propagation of sound on top of the fireball has been
discussed by J.Casalderrey-Solana and one of us in
\cite{CasalderreySolana:2005rf}. In that paper the fireball
expansion was modelled by the Big-Bang-like overall expansion of
the space, with the same  Friedman-Robertson-Walker metric as used
for cosmology. The focus of that paper was the effect of
time-dependent sound velocity, especially if the phase transition
is 1st order and it can vanish at some interval of $T$. The
interesting finding was a creation of the secondary -- and
convergent -- sound waves. This idea was further discussed in
\cite{Shuryak:2009cy} in connection with the ``soft ridge" issue,
but with the conclusion that if the  current lattice data on the
speed of sound is correct, the effect of the reflected wave is too
small.

\begin{figure}[h]
\begin{center}
\includegraphics*[width=7.cm]{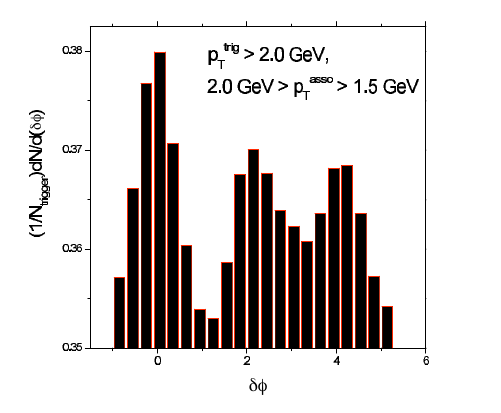}
\caption{(from   \cite{Andrade:2009em}) the two-particle
correlation function from the two-peak events, for particles with
transverse momenta indicated on the figure. }
\label{fig_threepeaks_corr}
\end{center}
\end{figure}

    In the same paper  \cite{Shuryak:2009cy}  it has been found that
the usual (unreflected) sound propagation should produce
characteristic  ``two-peak events", with the angle between  the
peaks reflecting the sound horizon and numerically being about 1
rad, see the blue curve in Fig. \ref{fig_twopeaks}(a). Andrade et
al \cite{Andrade:2009em} have independently come to the same
conclusion, see Fig. \ref{fig_twopeaks}(b) taken from their paper.
Note that their peaks are at the same angle: they are more
pronounced simply because the particle $p_t$ for which it is
plotted is higher.Note also a dip near zero, indicating that
nothing remains at the original location of the bump.
Andrade et al have further pointed out that the $two$-peak events
lead to a $three$-peak correlation function  shown in  Fig.
\ref{fig_threepeaks_corr}, with the side peaks now at twice larger
angle $\pm 2$ rad, now on the ``away" side from the peak.  (Note
that the central peak is about twice larger than the peripheral
ones: this is because it corresponds to events 11 and 22 while the
other peaks are 12 and 21 combinations, from the peaks 1 and 2 in
the two-peak distribution. )

 This observation  explained  what has been found earlier, in the ``event-by-event" hydrodynamical studies by
the Brazilian group (see \cite{Takahashi:2009na} and references
therein).
 The Brazilian group has used the initial condition from string-based model developed by
Werner  and collaborators. Recently this group also studied
event-by-event hydrodynamics \cite{Werner:2010aa} and claimed that
their correlation function describes the data on two-particle
correlations. They trace their origin to multiple ``bumps" in the
initial distribution that lead to the development of what they
called ``fingers". There are about $O(10)$ of them per event, each
corresponding to rather narrow angles. Presumably the
``event-by-event" complicated pictures are more or less  linear
superpositions of the independent two-peak structures from each
bump discussed above, because the sound waves (as $any$ Goldstone
modes) hardly interact with each other.  We will discuss all those
issues in detail: and for now let us only note that all these
works have  $not$ yet included the key phenomenon to be studied,
namely $viscosity$. It is easy to estimate that even the minimal
conjecture viscosity $\eta/s=1/4\pi$ would significantly modify
the amplitudes of those fluctuations.

The most striking outcome of all these works
\cite{Takahashi:2009na,Werner:2010aa,Alver:2010gr} is that they have
obtained  the  ``Mach-cone-like" correlation functions, with the
away-side ridges at $\pm 2 rad$ (this in the other hemisphere from the trigger) without any jets involved!
This created large enthusiasm, and attempts to disregard hard collisions and jets, even for the events with trigger hadrons of few GeV $p_t$.
However, there is no doubt that such particles would not be produced from hydrodynamics and are still due to
hard collisions producing jets.

  Although this work is in many respects a continuation of the paper \cite{Shuryak:2009cy},
we will focus on sounds and postpone the development of another
important idea proposed  there. In spite of its phenomenological
success, there are good reasons to think that the hydrodynamical
description of the RHIC evolution may be incomplete because in the
so called M-phase of the collision electric gauge fields should
remain unscreened for the time of evolution. It has thus been
proposed to upgrade hydrodynamics to (dual)magnetohydrodynamics,
including these unscreened electric fields in the stress tensor.
This changes the nature of the propagating modes, and it also
makes  possible the survival of the ``flux tubes" in the ``QGP
corona". We hope to return to those effects later.

\section{Setting up the problem}
\subsection{The main scales of the problem}

Before going to specifics, let us formulate the problem in a more
general form, which in a way connects it to the Big Bang
fluctuations.

Two generic scales are (i) the macroscopic  scale $R$ and the
microscopic scale $l$, being in the relation \be l \ll R \ee which
ensures such macroscopic tools as thermo and hydrodynamics to
work.

The macroscopic  scale $R$  is the size of the fireball in heavy
ion collisions and the curvature scale $a$ in the Big bang. Note
that both are in principle time dependent, demonstrating the
expansion of the system. However, while $a(t)$ changes by many
orders of magnitude, the fireball size increases rather modestly,
e.g. from 6 to 8-9 fm at its maximal size, for AuAu collisions at
full RHIC energy.

The microscopic scale $l$ is the mean free path for weakly coupled
systems (weakly coupled QGP or hadronic gas): in the case of
strongly coupled QGP (sQGP) it is just the inverse temperature
$l=1/T$.  For AuAu collisions at full RHIC energy $l$  changes
from .5 to about 1 fm, from initial to hadronization time. Thus
the large parameter  $R/l =O(10)$ in the region we use
hydrodynamics.

Now let us define two new scales. The first is the {\em sound
horizon} \be H_s=\int_0^{\tau_f} d\tau c_s(\tau)
\label{eqn_sound_horizon} \ee where the integral is taken from the
formation to freezeout time. At the freezeout the waves just stop
where they are, and the matter is split into independent
particles.

It is the same idea as suggested by Sunyaev and Zeldovich for the
Early Universe \cite{Sunyaev:1980vz}: the initial perturbation
(say higher density at some point) creates a sphere of such
radius, at which the density is a bit higher than the average:
when galaxies are formed they are correlated with that sphere and
thus is observed today in their correlation function. It turns out
that in the Big Bang this produces a ``standard ruler", which is
today of about  $H_s\approx$ 150 Mps, observed in galaxy's
distribution and in CMB correlations.

One of the main issues discussed in this paper is whether any
manifestation of the sound horizon scale can be observed in the
Little Bang. For example, one may think of angular correlations
with angles \be \Delta \phi \approx {2 H_s \over R }\ee or angular
harmonics with $m\sim 1/\Delta \phi $. In the cosmology such
angular momentum is $l\sim 200$. Going ahead of ourselves, we will
show that in our problem of the Little bang, for AuAu collisions
at RHIC, we will deal with $m\sim 3$.

The second scale is not important in cosmology and we would like
to call it {\em ``the viscous horizon scale"} $R_v$. Its verbal
definition is that it separates the wavelengths of the sound which
{\em are and are not} dissipated  by the  viscosity effects. The
smooth fireball and fluctuations are described by
\be %
T_{\mu\nu}=\tilde T_{\mu\nu}+ \delta T_{\mu\nu}
\ee %
The textbook dispersion law for the sound, including the viscosity
term,  is \be \omega =c_s k- {i\over 2} {4\eta \over 3 s } {k^2
\over T} \ee

After that Fourier transform puts it into momentum form, after
which one can solve the time dependence using the
momentum-dependent dispersion relation as well as the imaginary
part induced by viscosity.

(One may add bulk viscosity to this expression as well, but we keep the  shear  viscosity for now, assuming it is dominant.)

\be %
\delta T_{\mu\nu} (t) = exp\left(-{2 \over 3} {\eta \over s} {k^2
t \over  T } \right)  \delta T_{\mu\nu} (0)
\label{eqn_visc_filter}
\ee %

\begin{figure}[t]
\begin{center}
\includegraphics*[width=6.5cm,height=8cm]{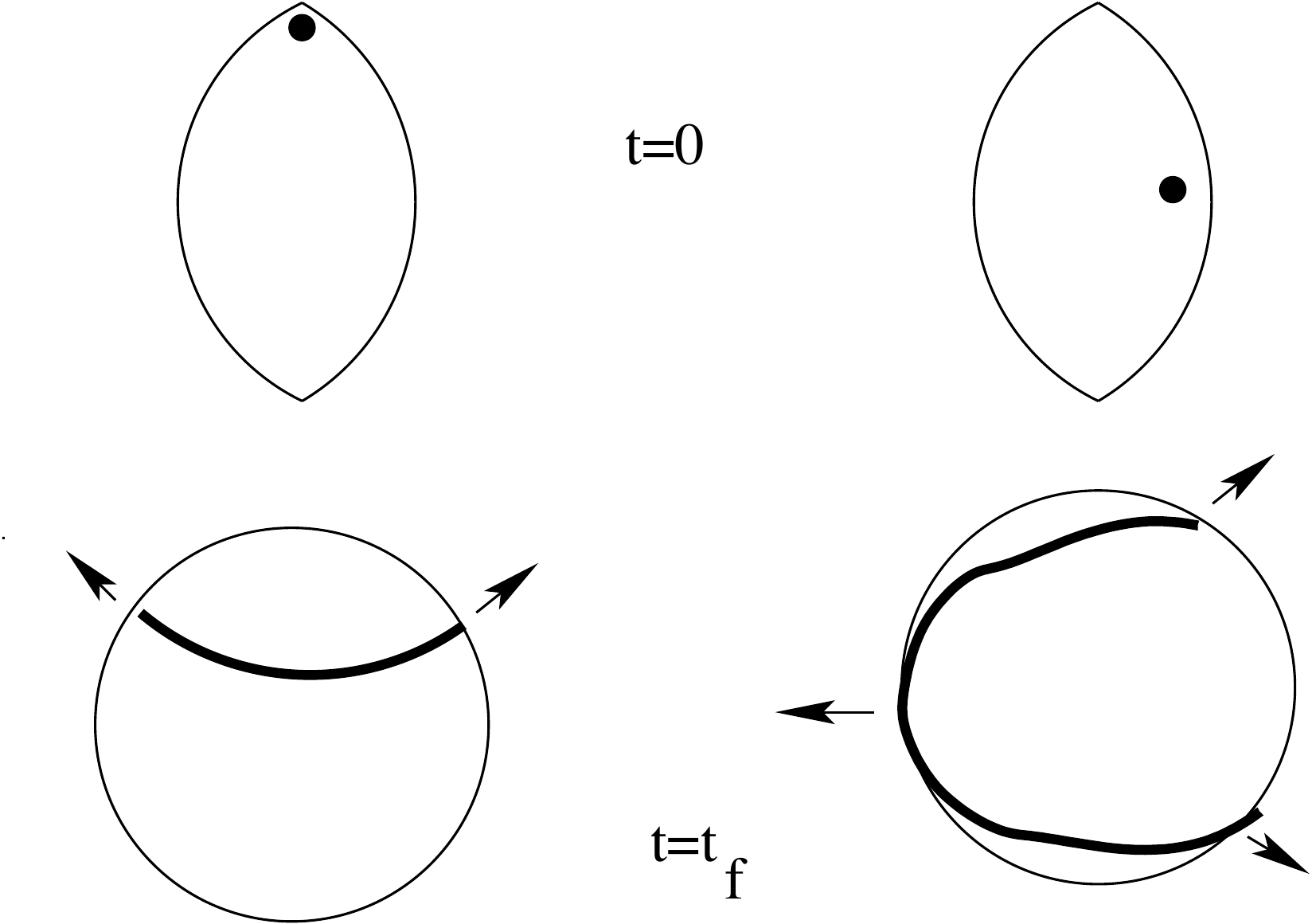}
\caption{Two upper picture correspond to initial time $t=0$: the system has almond shape and contains perturbations (black spots).
Two lower pictures show schematically location and diffuseness of the sound fronts at the freezeout time $t_f$. The arrows indicate
the angular direction of the maxima in the angular distributions, 2 and 3 respectively.
} \label{fig_23peaks}
\end{center}
\end{figure}

The spectrum of the original $t=0$ perturbations have harmonics of
the so called ``saturation scale" $Q_s$, which is for RHIC of the
order of   $Q_s\sim 1 GeV$. Even if one takes the minimal
viscosity $\eta/s=1/4\pi$, by freezeout $t\sim 10 \, fm/c$ this
exponent gets very large, damping such fluctuations to un
observably small magnitude. Only the harmonics below the new {\em
viscous survival scale} $k<k_v$ would survive, which  is
determined from the condition that the exponent above is less than
1
\be %
k_v ={2\pi \over R_v}=  \sqrt{ { 3T s \over  2 \tau_f \eta }} \sim
200  MeV \label{eqn_visc_horizon}
\ee %
(the number comes from an estimate $\eta/s=1/4\pi, T\sim 200 \,
MeV, \tau_f\sim 10 fm/c$).

What it means is that sound perturbation
 spheres (cylinders, cones etc) would not have the width of the original fluctuations, but they get significantly widened, with the width  of
the order of $1/k_v$. Note that while the radius of the spheres
increases linearly with time $\sim t$, this width increases only as
$t^{1/2}$, which means that although the spheres become more diffuse, they are also relatively sharper and sharper
as time goes by.

Let us finish this section by pointing out the main aims of this
investigation. By observing propagating sound perturbations one
would like to measure the two scales, $H_s$ and $1/k_v$,
experimentally, defining two key hydrodynamical parameters -- the
{\em speed of sound} and {\em viscosity}. The way to do so is to
change the geometry of the collision (by centrality) and the size
of the nucleus (by changing the beam A), and by observing all the
harmonics of the flow. The amplitudes of the higher harmonics,
dampened by viscosity, if measured, would provide an independent
measure of the viscosity.

 For central AuAu collisions at RHIC the hierarchy relation
between all those four scales  is  \be R  > H_s
> R_v  > l \ee
As some representative numbers let us mention $8,4,2,0.3 \, fm$,
respectively. The observation angle of the ``peaks" is \be \Delta
\phi\approx 2 {H_s \over R}\ee Their angular width is   \be \delta
\phi=R_v/R \ee so harmonics larger than $m> R/R_v$ can hardly be
found, as they are dissipated.

However for mid-central collisions the width (short size) of the
``almond" $R_x$ becomes comparable with $H_s$, and for more
peripheral ones $R_x<H_s$. As a result, one expects the sound to
traverse the whole fireball and deposit some amount of (entropy)
density to its side opposite to the original fluctuation, as it
speeded up with the flow. In this case one expects 3-peak events:
see Fig.\ref{fig_23peaks} for explanation.

 So, one of our suggestions to experimentalists is to locate the centrality at which such transition occurs.

\section{The initial state fluctuations}
\subsection{Generalities}
    Let us start with a comment on what we would call the ``initial
state". This term is currently used in at least three different
settings: \\ (i) The wave function of the colliding nuclei,
expressed either in terms on the nucleons (their positions in the
transverse plane just prior to the collisions) or in terms of
partonic degrees of freedom (positions and longitudinal momenta).
Another version of it is the ``Color Glass Condensate" (CGC) described as an ensemble of classical gauge fields.\\
(ii) The state just after the (Lorentz contracted) nuclei passed
each other. It is either the partonic state, including partons
newly produced
in a collision; or the so called GLASMA, in the classical field  description. \\
(iii) The state after approximate equilibration is reached, so
that macroscopic (hydrodynamical) description can be started.

It is the last one which we mean in this work, as we would apply
hydrodynamics as a tool, translating properties of the initial
conditions into the final state observed in the experiment.
Therefore our ``initial state" should correspond to about one unit
of the relaxation time after the actual collision, or numerically
at a proper time of the order of 1/2 fm/c.  Thus the inhomogeneity
of the initial wave functions should be already smoother than at
time zero, by this (so far poorly understood) relaxation process.

As we detail below, this state will be described by some
``average" or zeroth-order shape of the fireball (depending of
course on the impact parameter, the colliding nucleus and the
collision energy), plus ``fluctuations" characterized in the first
order by an ensemble of small perturbations of the average shape
described by Fourier coefficients and phases $\{ \epsilon_n,\psi_n
\}$. Generic expressions would include the zeroth order
ensemble-average deformations $<\epsilon_n>$ and deviations which
have no average but fluctuations $\delta\epsilon_n^2
=<\epsilon_n^2>-  <\epsilon_n>^2$.

    The simplest situation, happening for the second harmonics and
sufficiently peripheral collisions, is that the average is much
larger than the fluctuations, $<\epsilon_2> \gg \delta\epsilon_2$.
If so, one may assume  Gaussian form of the fluctuations with the
width given by $\delta\epsilon_2$. But the situation is quite
different for near-central collisions, for which both terms in
$\epsilon_n$ come from fluctuations. Their distribution are
obviously non-Gaussian because they are all positive by
construction. So one must  introduce and study their higher powers
or cross-correlations (such as $<\delta \epsilon_{n_1}  \delta
\epsilon_{n_2}  \delta \epsilon_{n_3}  >$ and their
generalizations with certain combination of phases (see below).
All of those should in principle be provided by the ``initial
state models", of which we select Glauber model as the simplest
example.

The separation of the initial state fluctuations from all other
fluctuations (e.g. fluctuations during the hydrodynamical
evolution, hadronization and the freezeout) is possible because of
the fundamentally different number of relevant degrees of freedom
defining their magnitude. As we will detail in the next section,
the so called Glauber fluctuations due to various number of
``wounded" (or participant or interacting) nucleons   are of the
order of \be  \epsilon_n \sim {1 \over \sqrt{N_p} }\ee where the
number of the participant nucleons $N_p\sim O(100)$, being limited
from above by the total nucleon number $2A\sim 400$.

Further fluctuations are determined similarly, but with the number
of participants $N_p$ substituted by the {\em much larger} number
of partons involved, or the total multiplicity $N_{hadrons} \sim
10^4$ (for RHIC and LHC it is factor 2 different). That is why one
may, to certain accuracy, ignore all later-time fluctuations and
assume that observable fluctuations in particle spectra and
correlation functions are one-to-one translated from the initial
state ensemble. Thus we use hydrodynamical equations as a fully
deterministic tool,  by itself producing no random numbers at all.

Furthermore, for near-central collisions all  $\delta\epsilon_n$
are small, of the order of several percents. So, independently of
their possibly complicated distributions and cross correlations,
the  hydrodynamics  applied in  linear approximation should be
quite reliable tool. Thus hydro equations can be linearized and
the linear response coefficients $\delta v_n/\delta\epsilon_n$
calculated. If so,  it does not matter what  the actual magnitude
of the deformation $ \delta\epsilon_n$ is. Also the linearized
perturbations  do not interact with each other.

  Although we will focus on those calculations in our next paper,
let us note here two things. One simple fact is that while angles
$\psi_n$ of the fireball deformations indicate the $maxima$ of the
distribution (the corners of triangle, square and other polygons),
hydro flow goes along their $sides$. Therefore the observed flow
angles $\xi_n$ are rotated from the deformation angle as follows
\be \xi_n = \psi_n+ {\pi \over n} \label{angle_shift}  \ee Our
second comment is that higher harmonics $n$ are supposed to become
oscillatory in time, displaying acoustic sound properties. Naively
one may think that at freezeout this leads to their random phases,
and thus those can be ignored. This is however not true, as only
the coherent sum of all harmonics with their correct phases will
reproduce a propagating sound wave. Finally, if needed,  nothing
prevents one from solving full nonlinear hydrodynamics (without
linearization). In fact it is done by a few groups devoted to
``event-to-event hydrodynamics". If  done one can  also calculate
the nonlinear effects, e.g. appearance of the 4-th harmonics $v_4$
in spectra coming not from $\delta \epsilon_4$ but from $(\delta
\epsilon_2)^2$ .

\subsection{Fluctuations in the Glauber model: the amplitudes }

    Our ``Glauber model" is a bit different from that used widely by
experimentalists. Both assume that  initial state fluctuations
originate from the nuclear wave functions. The ``usual Glauber"
uses   randomly placed coordinates of the individual nucleons in
the nuclear wave function. However, the nucleons themselves are
complicated objects and their interactions are also strongly
fluctuating: since there are studies of that we decided to include
this source of fluctuations also. This changes numbers a bit, but
was found not to be important for any of the qualitative
conclusions to be reached.

   The nucleon  fluctuations we  included  via  the {\em
fluctuating NN cross sections} which are to a certain degree known
and studied via diffraction, see   \cite{Baym:1995cz} for the
details and earlier references.  Naively, from the well known fact
of a nucleon being made of quite a large number of partons one
might conclude  that those fluctuations are small,
$O(1/N_{partons})$:  but this is $not$  the case. In our
simulation we have assumed the cross section $\sigma_{NN}$ to be
the random Gaussian variable with the variance \be
w_{NN}={<\sigma^{2}_{NN}>- <\sigma_{NN}>^{2} \over
<\sigma_{NN}>^{2} } \approx 0.25 \ee

First, like in \cite{Alver:2010gr}, we simulate a large ensemble
of collisions and calculate the magnitude of the $\epsilon_{n}$
for several lowest harmonics (up to
 6).
 Their definition is via the Fourier expansion for a single particle distribution
\begin{eqnarray}
f(\phi) & = & \frac{1}{2\pi} \left(1 + 2\sum_n \epsilon_n
cos(n(\phi-\psi_n))\right)
\end{eqnarray}
where the $\epsilon_n$ are the participant anisotropies and the
$\psi_n$ are the angles between the x axis and the mayor axis of
the participant distribution.

The participant anisotropies are calculated from
\begin{eqnarray}
\epsilon_n & = &
\frac{\sqrt{\left<r^n\cos{(n\phi)}\right>^2+\left<r^n\cos{(n\phi)}\right>^2}}{\left<r^n\right>}
\end{eqnarray}
This expression is calculated in the center of mass of the
participant nucleons for each event. Therefore the dipole moment  $n=1$ made out of the average coordinates
\be <x>=\left<r\cos{(\phi)}\right>=0, \,\,\,\,\  <y>=\left<r\sin{(\phi)}\right> =0\ee are
zero by definition.

The 2-d shape of the event can in principle be expanded in the
double Taylor series in $x,y$ or in double series over moments
$r^m cos(n\phi),r^m sin(n\phi)$ with integer $m,n$. An even better
definition would be to follow the customary statistical trick and
write the distribution as the $exponent$ containing the
``generating function" of the angular dependence expended in
harmonics \be P= F_1(r) exp(F_2), F_2= \sum_{n>0} r^n \epsilon_n
cos(n(\phi-\psi_n) \ee In this way the positivity of the
distribution function, as well as inclusion of trivial higher
order effects are ensured.

\begin{figure}[t]
\begin{center}
\includegraphics[width=8 cm]{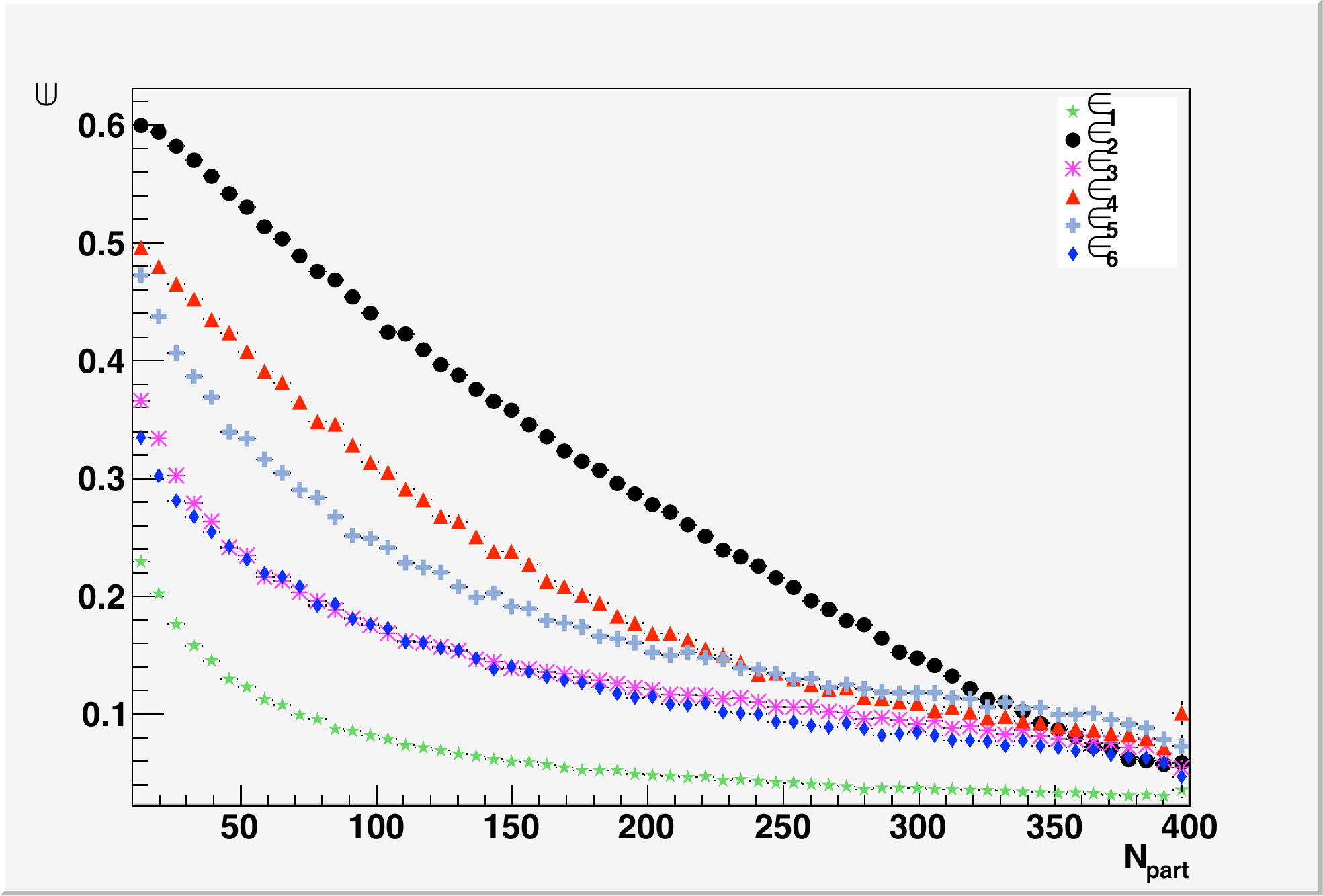}\\
\includegraphics[width=8 cm]{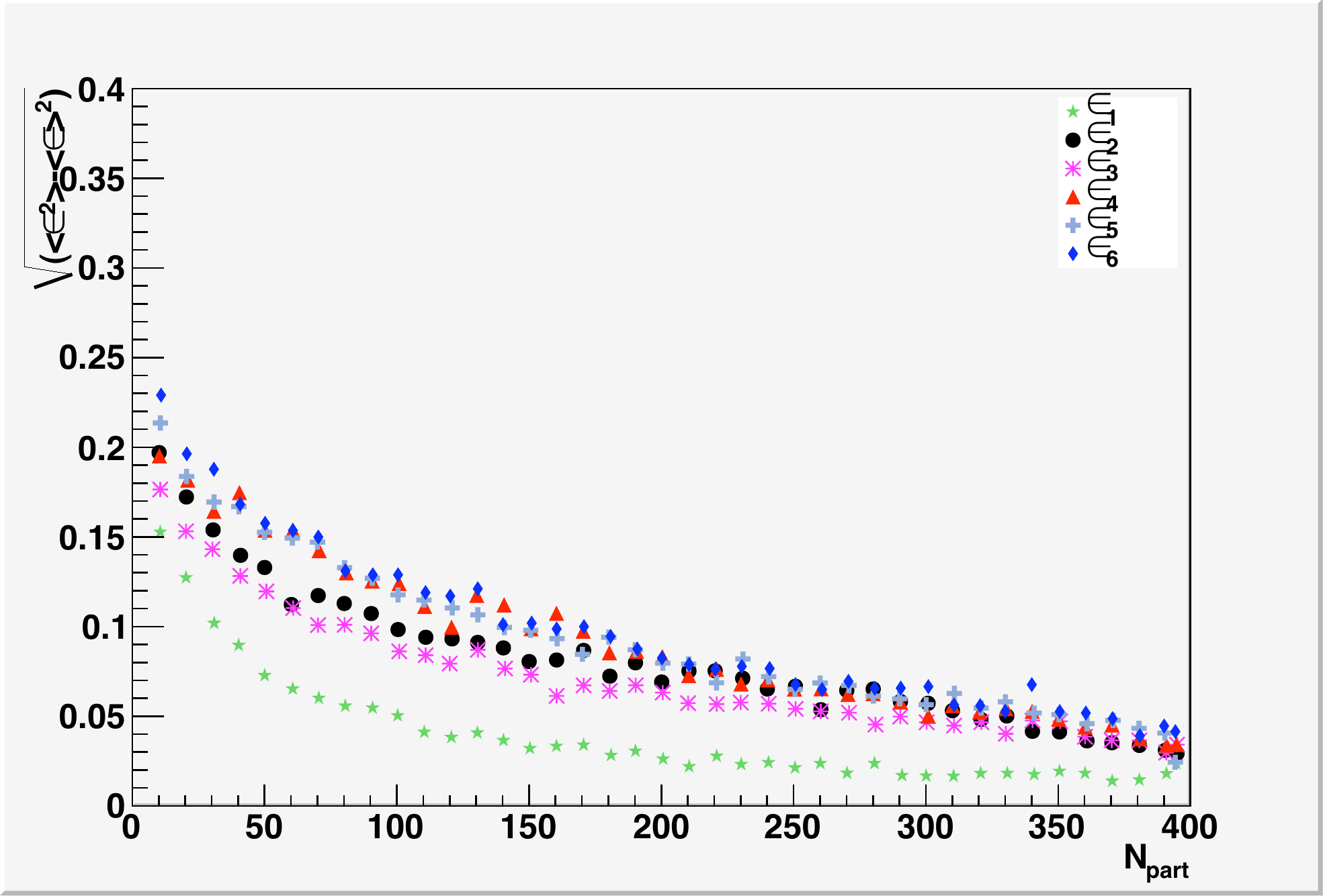}
\end{center}
\vspace{-5ex}\caption{Average anisotropies (upper plot) and their variations (lower), as a function of centrality expressed via the number of participants
$N_{part}$ }\label{anisotropy}
\end{figure}

Since the dipole $m=n=1$ term is zero by construction, we define
the first odd deformation $\epsilon_1, \psi_1$ using the  term of
the expansion $m=3,n=1$ which appears together with the triangular
 deformation $m=n=3$
\begin{eqnarray}
\epsilon_1 & = &
\frac{\sqrt{\left<r^3\cos{(\phi)}\right>^2+\left<r^3\cos{(\phi)}\right>^2}}{\left<r^3\right>}
\end{eqnarray}
The anisotropies calculated in this way are plotted in  figure
\ref{anisotropy} for $n=1,6$.
The plot shows that the eccentricity has the largest value for the
well known elliptic deformation $\epsilon_2$ and a nonzero value
of triangularity $\epsilon_3$, in agreement with the results
reported in \cite{Alver:2010gr}.  Note that for the near-central
collisions $N_{part}>300$ the elliptic deformation is no longer
dominant, and it is also due to fluctuations. This conclusion
becomes evident as one looks at the lower plot in
Fig.\ref{anisotropy}, which shows the variations of these
$\epsilon_n$.

One observation coming from these results is that all other
deformations (except for $\epsilon_1$, small because the ``true
dipole" remains zero)  are all comparable, ranging from $O(1/10)$
for central collisions to 0.3 -0.5 for most peripheral ones. While
in the central bins these perturbations can be considered small
and treated as Gaussian random variables, it is clear that for
most peripheral bins (when the number of participants is  smaller)
the fluctuations are large and thus must be non-Gaussian.

Another consequence is that there is absolutely no ground to
single out $\epsilon_3$: in fact both $\epsilon_4$ and
$\epsilon_5$ are larger that $\epsilon_3$ and $\epsilon_6$ is
about of the same order as $\epsilon_3$.

The last point is that their variations (the lower plot) are all
comparable to the magnitude.  Yet the definition of deformations
are such that they are always positive, for each event. This is
one more reason that the amplitudes cannot have Gaussian
distribution, deviating from it at least for the smallest values.

  \subsection{Fluctuations in the Glauber model: the angles }

\begin{figure}[t]
\begin{center}
\includegraphics[width=8 cm]{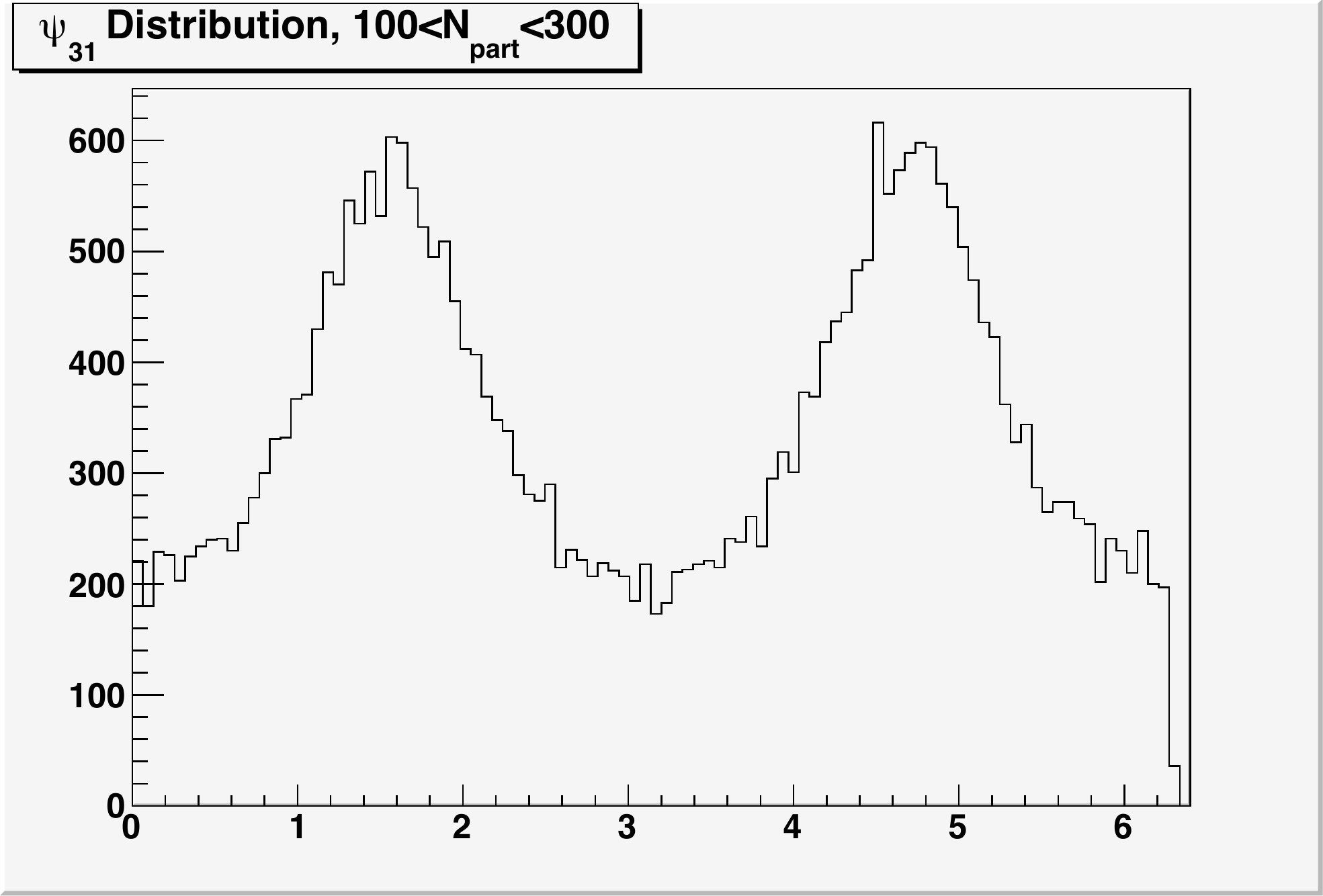}\\
\includegraphics[width=8 cm]{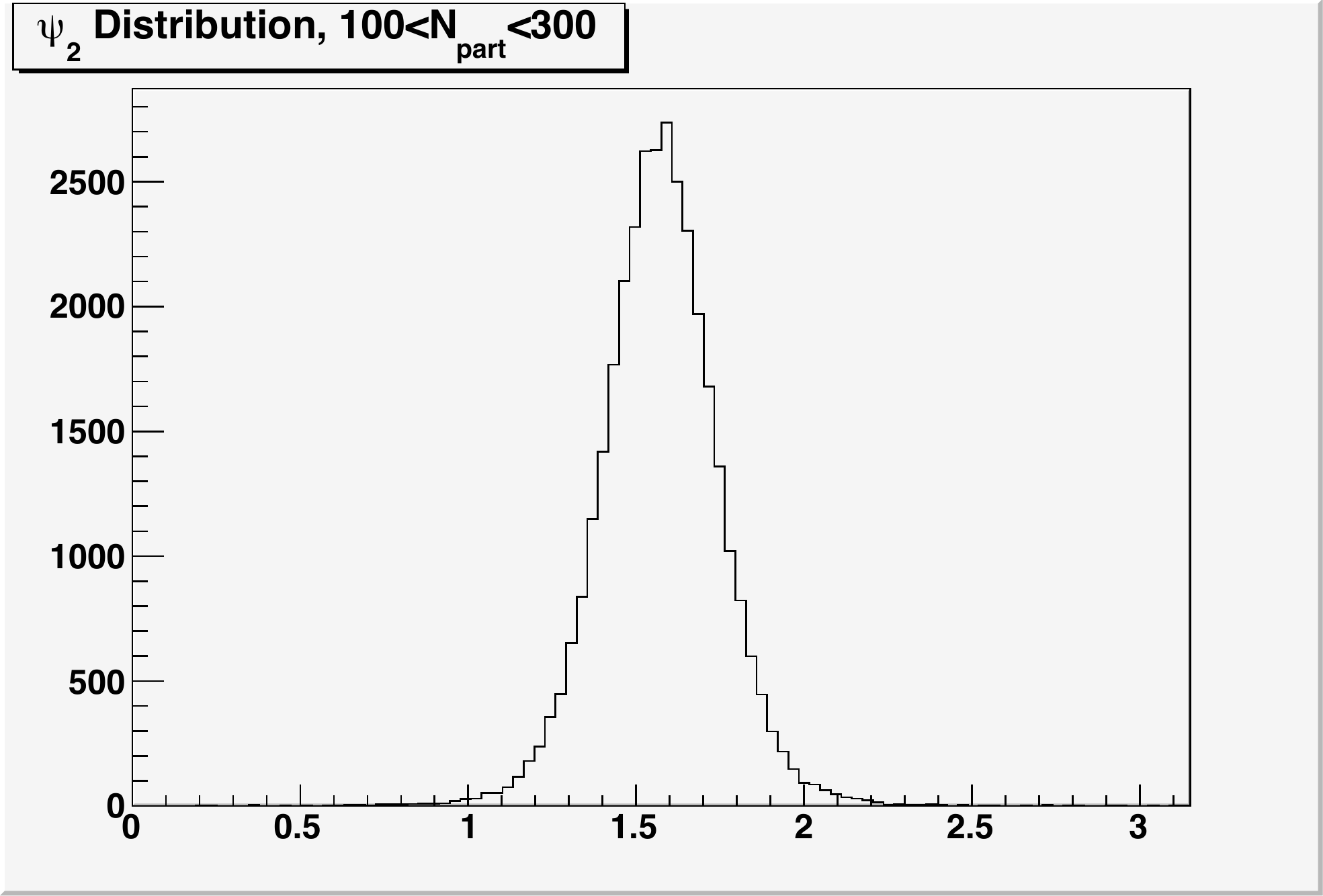}\\
\includegraphics[width=8 cm]{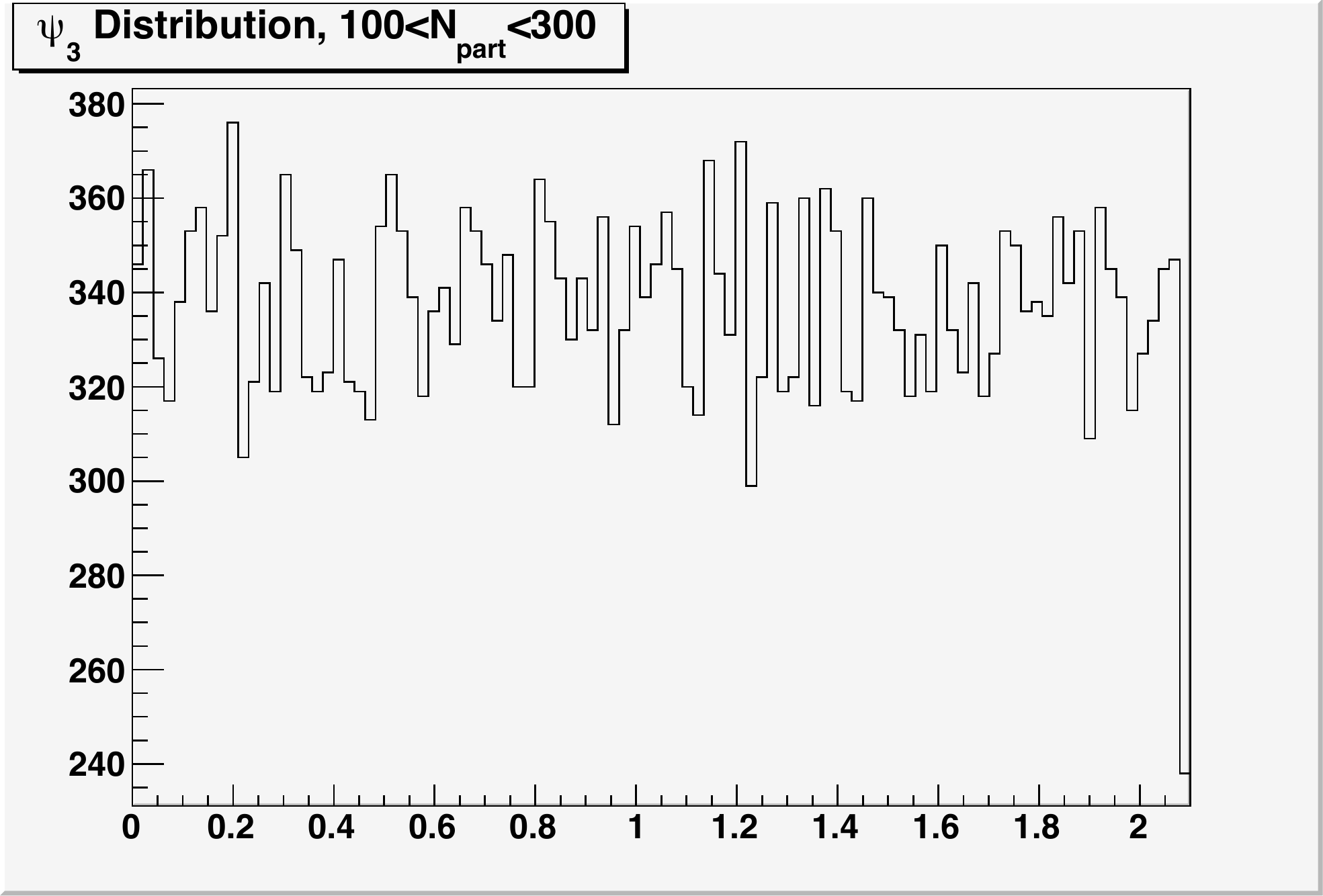}\\
\end{center}
\vspace{-5ex}\caption{Distribution of the angles $\psi_{n}$ for the first three
harmonics, the centrality bin used is  $100<N_{part}<300$}\label{angdist13}
\end{figure}

\begin{figure}[t]
\begin{center}
\includegraphics[width=8 cm]{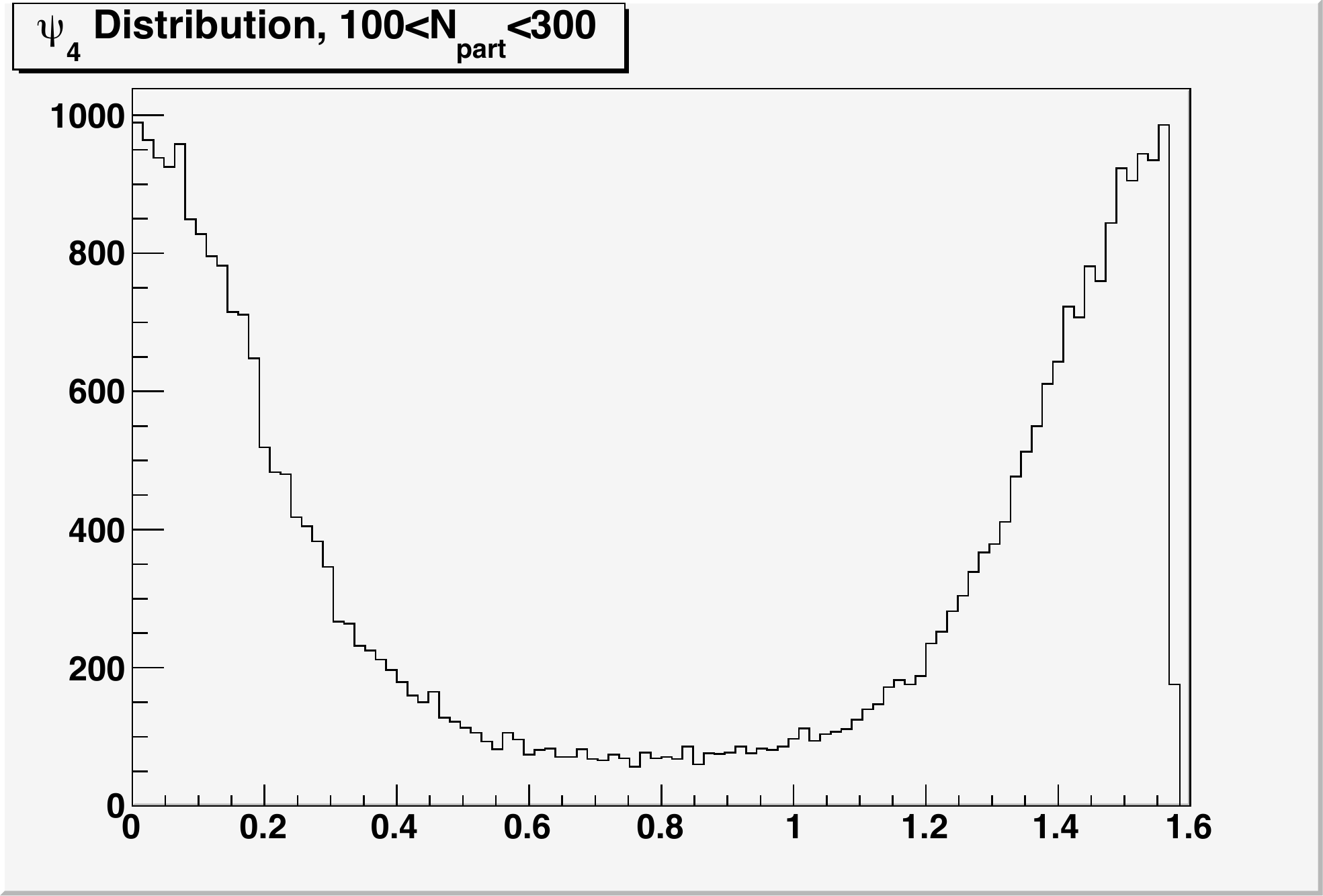}\\
\includegraphics[width=8 cm]{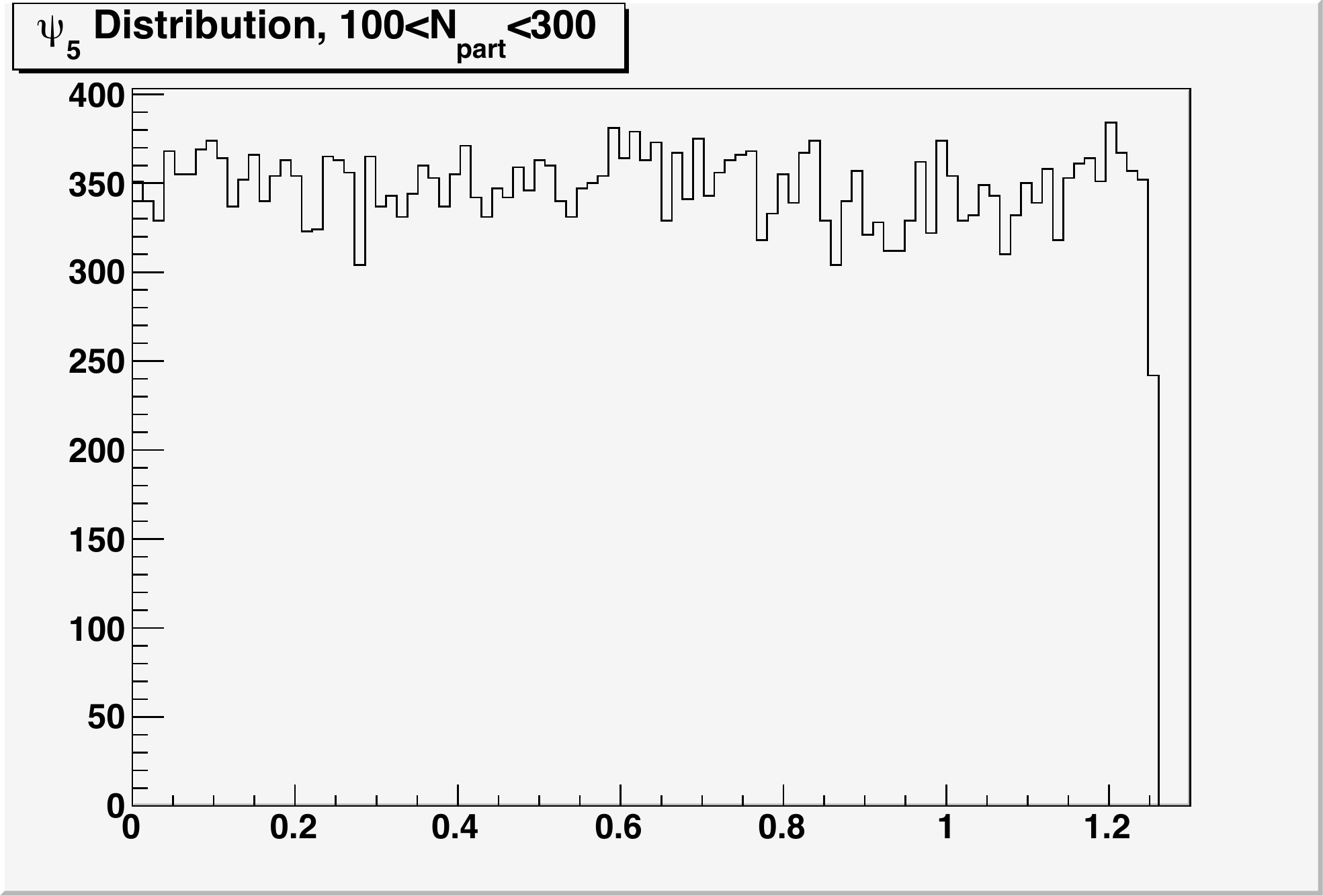}\\
\includegraphics[width=8 cm]{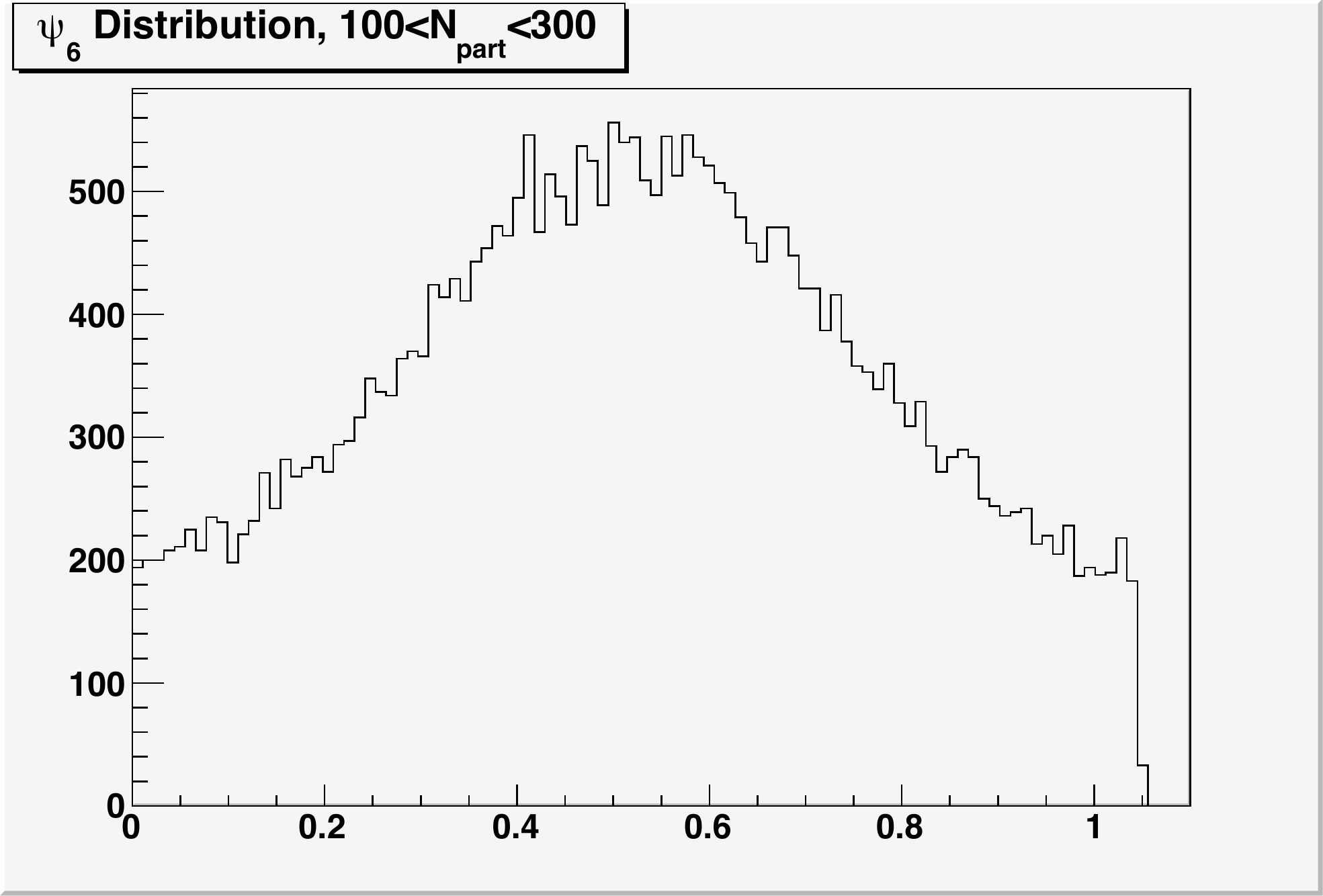}
\end{center}
\vspace{-5ex}\caption{Distribution of$\psi_{n}$ for the harmonics 4-6, same centrality }\label{angdist46}
\end{figure}

The angles
$\psi_n$ are defined by:
\begin{eqnarray}
\tan{\left(n\psi_n\right)} & = &
\frac{\left<r^n\sin{(n\phi)}\right>}
{\left<r^n\cos{(n\phi)}\right>}
\end{eqnarray}
and to calculate  $\psi_1$  we use:
\begin{eqnarray}
\tan{\left(\psi_1\right)} & = &
\frac{\left<r^3\sin{(\phi)}\right>} {\left<r^3\cos{(\phi)}\right>}
\end{eqnarray}
Using these expressions we obtain the distribution of the
$\psi_n$'s for the first six harmonics as shown in
Figs \ref{angdist13},  \ref{angdist46}. 
In order to better understand the behavior of these angles we will
now study their correlation.

 (Note that our angle definition is different from the one by Alver-Roland \cite{Alver:2010gr}: we do not  include extra phase $\pi/n$ between the flow and deformation directions, see
 below.)

 Let us comment on these distributions, starting from the even ones.

 The most obvious one is a distribution of the second (elliptic)
harmonic: as seen in Fig.\ref{angdist13} the angle $\psi_2$ is
strongly peaked at $\pi/2$, corresponding to an elongation of the
system in the $y$ direction, as of course one expects from the
overlap ``almond" of two nuclei. The distribution of the 4-th
angle $\psi_4$  in Fig.\ref{angdist46} shows peaks at angles 0 and
$\pi/2$: but since quartic symmetry of the 4-th harmonics it
simply means that the maxima of the distribution tend to be
aligned with the coordinate axes x and y. The distribution of the
6-th harmonics is different: it is peaked at the angle  $\pi/6$.
This means that it has no maximum at $x$ direction but rather in
$y$. In conclusion, all even harmonics are strongly correlated
with the reaction plane, all of them producing maxima along the
$y$ (out-of-plane) direction.

The distribution of the angle $\psi_1$ is nonzero at all angles,
which means it is not very strongly correlated with the reaction
plane. It has two maxima, at $\pm y$ directions, to be called
``tip" fluctuations. Although the contribution from angles $0,\pi$
or x-directions is about twice smaller, it  also makes an
important contribution: we will call it ``waist" fluctuations.
Note that while the area of the ``waist" is larger than ``tips":
and yet its contribution is smaller.

\begin{figure}[t]
\begin{center}
\includegraphics[width=9 cm]{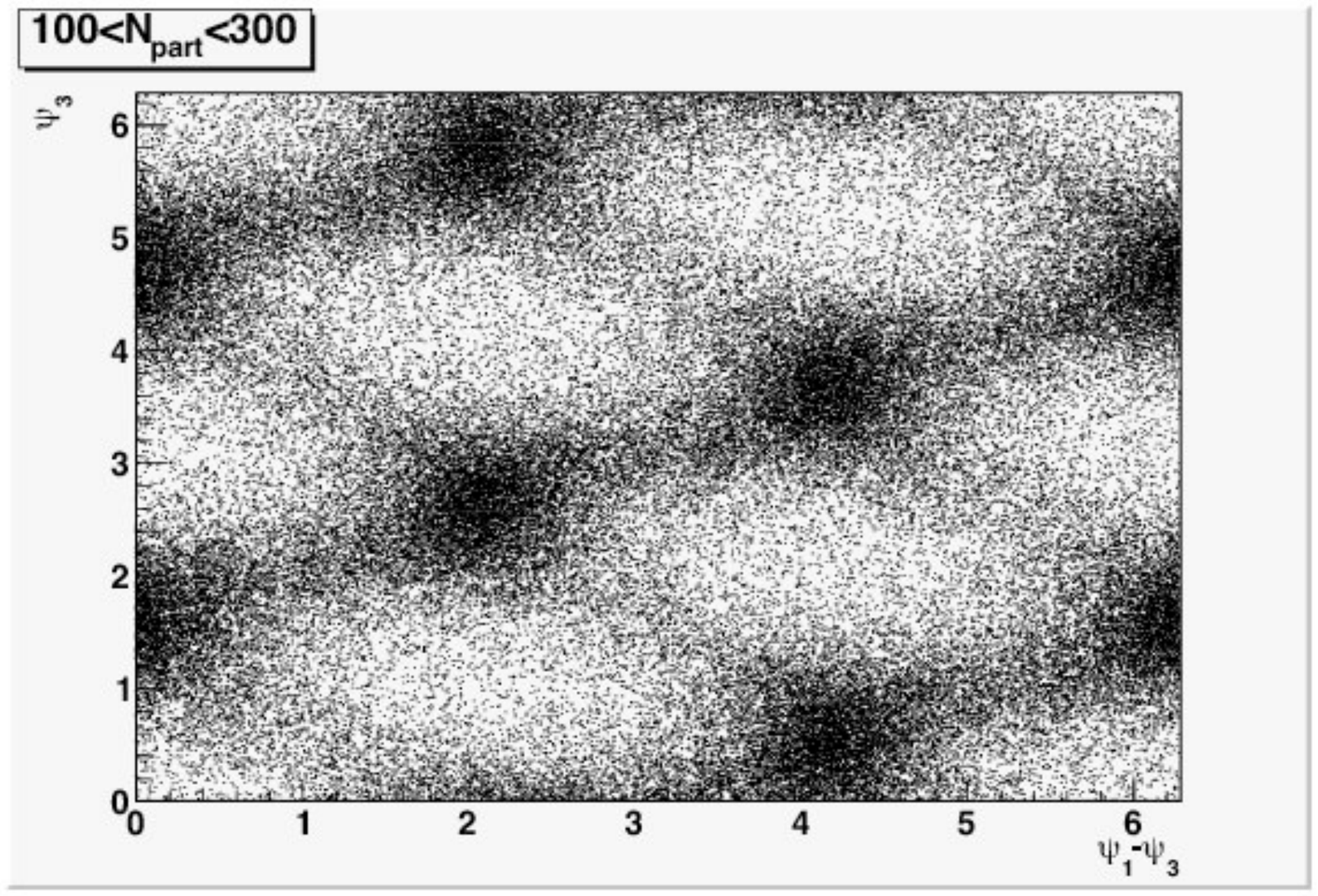}\\
\includegraphics[width=9 cm]{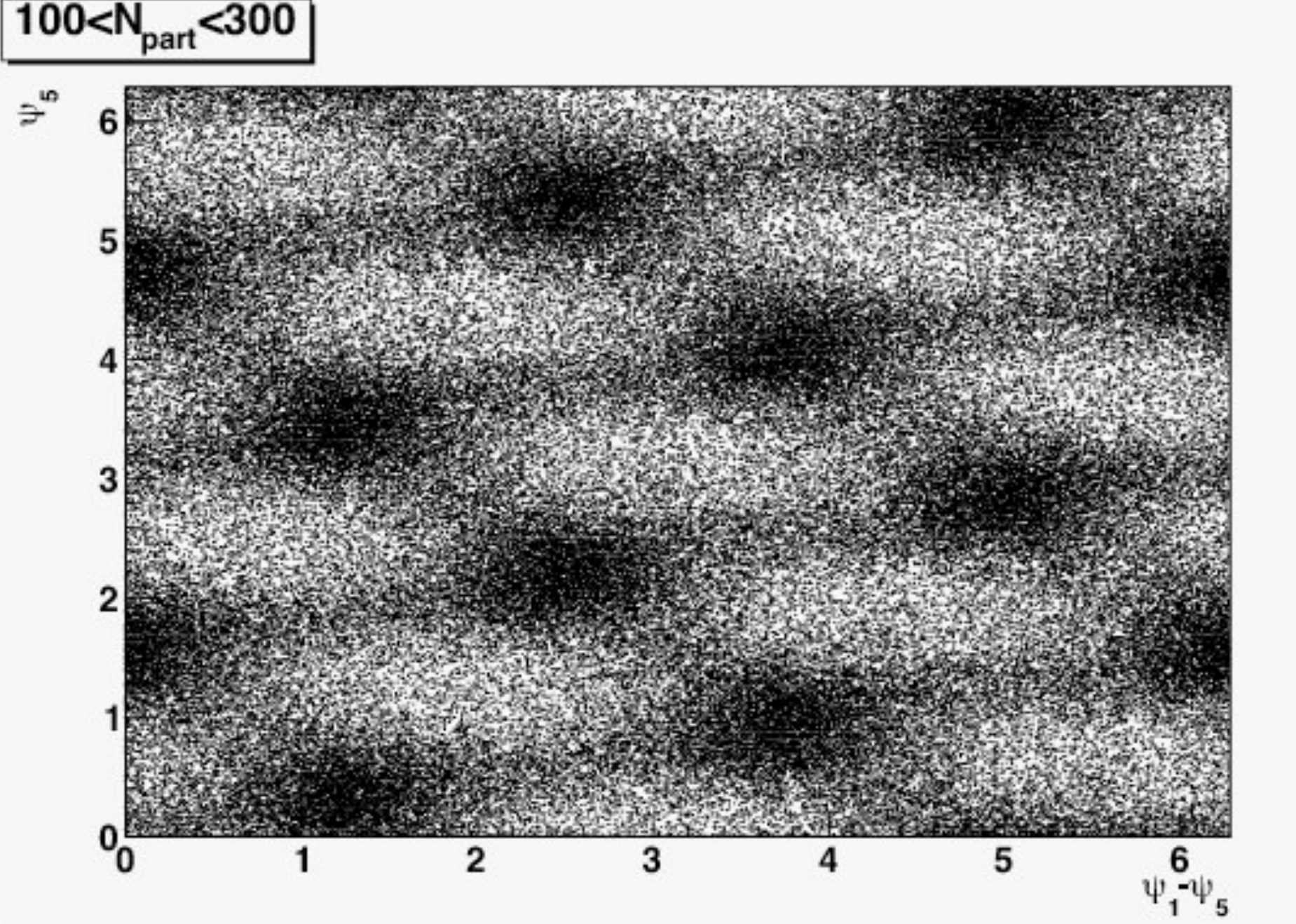}
\end{center}
\vspace{-5ex}\caption{Scatter plot of the $\psi_{3}$ vs $\psi_{3}-\psi_1$ (above), and of the $\psi_{5}$ vs $\psi_{5}-\psi_1$ (below),
the same centrality }\label{diff}
\end{figure}

 The distribution over $\psi_3,\psi_5$  in these figures looks
completely uncorrelated with the reaction plane. (This fact has
also been noticed in \cite{Alver:2010gr} and by others.) However,
further scrutiny shows that they are in fact well correlated with
$\psi_1$, see Fig. \ref{diff} (in which we included points
repeated by periodicity). The distribution can be crudely
characterized by some ``bumps" plus ``stripes" connecting them.

 The interpretation of the  ``bumps" is that all of them correspond
to events with additional ``hot spot" at the ``tips" of the
almond. It is a very natural place for maximal fluctuations, for
two reasons. First, this is where the participant density in both
nuclei is near zero. Second, because of the factor $r^3$ they have
larger weight than  all other places.

 There are two kinds of ``stripes", with positive and negative
slope in  Fig. \ref{diff}. The latter one simply follow from
$\psi_1$ distribution, while the former one is indeed a nontrivial
correlation between the angles whose origin we cannot explain. We
will continue to discuss its manifestation a bit later. The
correlation of $\psi_5$ with $\psi_1$ is very similar. The
``bumps" at $\psi_5-\psi_1\approx 0$ again mean $\pm y$ the
direction or the ``tips". The plot has similar ``stripes".

Going a bit ahead of ourselves, let us study the ``resonant"
combinations of angles, as well as angles and amplitudes. As we
explain below, those particular combinations of the amplitudes and
phases or two harmonics are \be  <\epsilon_{n_1}  \epsilon_{n_2}
cos \left( n_1 \psi_{n_1}- n_2 \psi_{n_2} \right)>
\label{eqn_reson} \ee especially in the case when $n_1,n_2$ differ
by two units. We have studied two first examples of the kind, with
odd harmonics 1,3,5.

\begin{figure}[t]
\begin{center}
\includegraphics[width=9 cm]{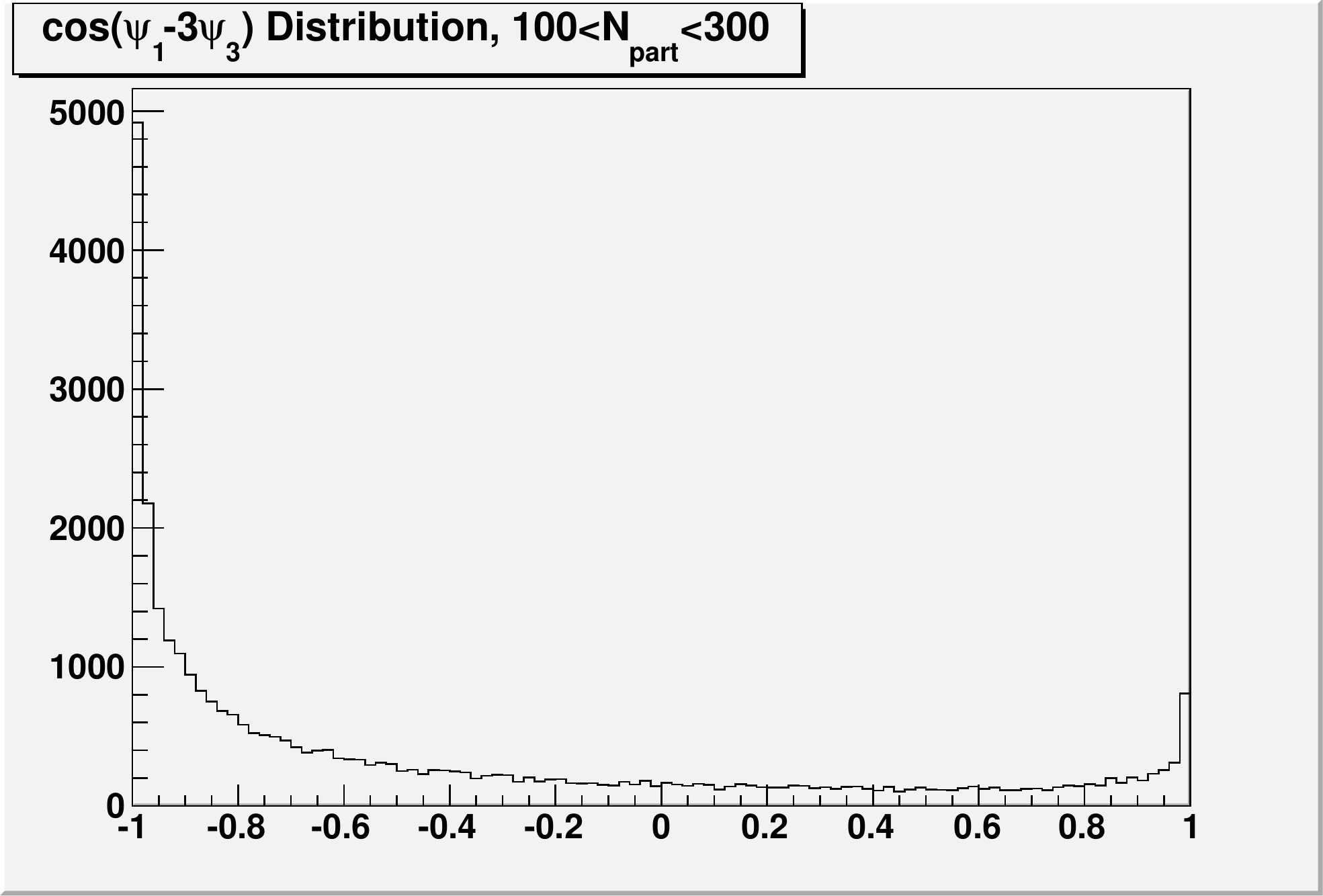}
\end{center}
\vspace{-5ex}\caption{Scatter plot of   $cos(3\psi_{3}-\psi_1)$ }\label{cos31}
\end{figure}

One interesting distribution, shown in Fig.\ref{cos31}, is that
over the $cos$ term itself, for the particular combination of the
1-3 phases. It consists of two clearly different parts: a very
narrow peak near $-1$ and wide flat distributions between -1 and
1. This plot demonstrates qualitative feature of the phase
distribution which was pointed out above.  One explanation of the
peak near -1 (the angle combination is $\pi$) comes from the
fluctuations at the ``tips" of the almond, when both $\psi_1$ and
$\psi_3$ are strongly correlated close to $\pi/2$. However, the
second interesting correlation seen as ``positive slope lines" in
Fig.\ref{diff}a   because for them $\psi_1-3\psi_3=\pi$ as well.
A similar situation happens for other odd harmonics.

\begin{figure}[t]
\begin{center}
\includegraphics[width=9 cm]{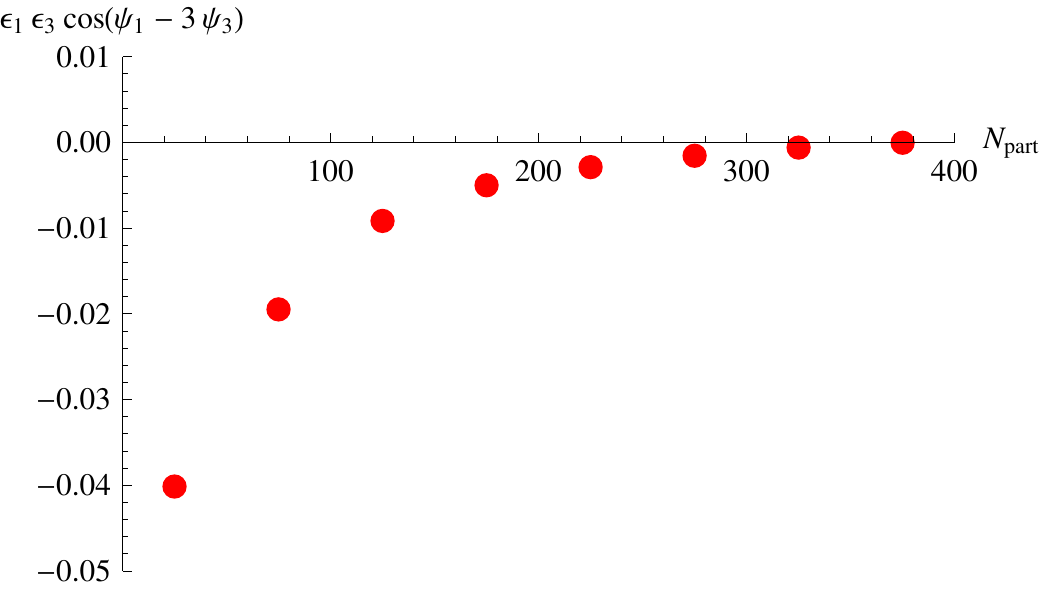}\\
\includegraphics[width=9 cm]{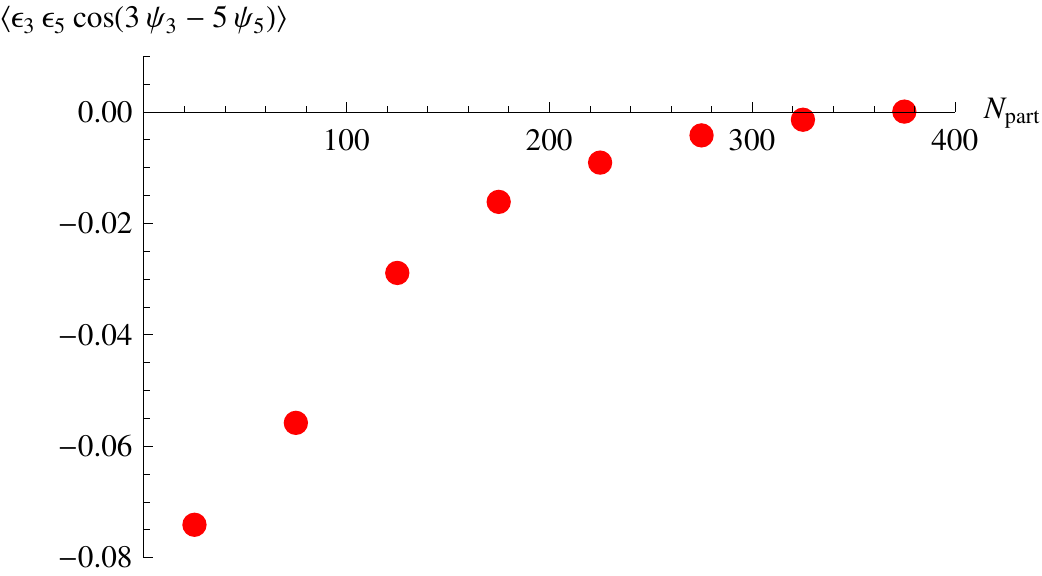}
\end{center}
\vspace{-5ex}\caption{Correlators $\langle\epsilon_{1}
\epsilon_{3} cos \left( \psi_{1}- 3 \psi_{3} \right)\rangle$ (top)
and $\langle\epsilon_{3} \epsilon_{5} cos \left( 3\psi_{3}- 5
\psi_{5} \right)\rangle$ (bottom)as a function of the number of
participants. The error bars are omitted since they are smaller
than the dots. }\label{combi}
\end{figure}

The average value of the combinations (\ref{eqn_reson}) for 1-3 and 3-5 harmonics as a function of centrality are shown in Fig.\ref{combi}.
All values are negative, as the sign is dominated by a peak in $cos$ near -1: the other component more or less averages out. We thus conclude that
experimental measurements of the amplitude of such correlations, with the magnitude and the sign, will be especially sensitive to
the ``almond tip" fluctuations.

Summarizing the observed pattern: we have found that all odd
angles are well correlated with each other, forming the ``stripes"
and ``bumps" shown in Fig.\ref{diff}.
The fluctuations and correlations seem to be stronger from the
``tips"of the almond.

\subsection{Comments on other initial state  fluctuations }
So far the only source of fluctuations included has been  (i)  the
coordinate part of the nuclear wave function prescribing the
nucleon positions in the transverse plane, and (ii) the
event-by-event fluctuation of the $NN$ cross section. We found
that the former effect dominates and the latter only provides
small corrections.

While other sources of fluctuations clearly are subject for future
studies, we still provide some comments on those.

One important type of  ``initial state" fluctuations is of course
hard parton scattering events, resulting in jet production. The
rate of those very strongly depend on the exact definition of the
cutoff beyond which the  momentum transfer involved is
characterized as ``hard". There are vastly different opinions on
where this boundary is theoretically, and experimentally it
depends on whether such events are triggered by single large-$p_t$
hadron or by some jet-finding algorithms. Jet production and
quenching is of course of high interest, but those should be
studied only in small fraction of all collisions selected by
separate triggers. For global fluctuations those can safely be
ignored, as the probability of ``hard" events is smaller than that
of the fluctuations we study.

In the Glauber approach that we used, the local density of
produced matter is assumed to be simply proportional to the local
density of all participant nucleons, $N_p(A_1)+N_p(A_2)$. However,
when this density is high enough, it has been argued that the so
called ``saturation" phenomenon should take place, because of
parton absorbtion processes in the wave function. Other
expressions for local matter density have been proposed, e.g.
$\sim min(N_p) ln(max(N_p)/min(N_p))$ by Kharzeev, Nardi,Levin,
where min and max refer to the smaller and the larger of the two.
Those are typically the amplitudes of the fluctuations.

A well known approach to their description is the so called
``glasma", which calculates those color fields from random color
charges of the leading (larger x) partons of  the two colliding
nuclei. Asymptotically (in a very large nuclei or at very high
energy) McLerran and Venugopalan \cite{McLerran:1993ni} argued
that   at a particular location in the transverse plane the color
charges of partons must be uncorrelated  because they come from
different nucleons.  Therefore it is usually assumed that their
color charges fluctuate as random variables. If so, the resulting
fluctuations are small, as the total number of partons is very
large.

Application of such ideas usually keep the average values of
those, such as $A^{1/3}$ or so. The point of our comment is a
warning, that such simplified ideas cannot be used for determining
the fluctuations. It has been known for a long time, that while at
very small x the partons, mostly gluons, become numerous, they are
still very tightly correlated in the transverse plane. The size of
the ``gluonic spot" in the nucleon has been known for long time
from diffractive form factors, and in more recent form, from HERA
photon diffraction into $J/\psi$. This spot is small, and
therefore bright. As documented e.g. in  \cite{Kowalski:2003hm}
(their Fig.23), the gluon density at the center of the nucleon is
about as high as in the center of $Ca_{40}$. Therefore, large
number of gluons does not yet imply that all of them merge in the
transverse plane into more or less homogeneous distribution: the
positions of the incoming nucleons still remain the dominant
source of the initial state fluctuations.

As the ``gluonic spots" from single nucleons remain the source of
initial state fluctuations, one may ask if numerous partons coming
from it may be  correlated in their quantum numbers, forming
specific large-amplitude color fields. One particular kind of such
fields got special attention: those are the  topologically
nontrivial gluonic configurations called {\em the QCD sphalerons}
\cite{sphalerons}. They are gluonic field configurations which
originate from excitations of the topologically nontrivial vacuum
fields (instantons). While they rapidly explode into multiple
gluon state,  they  strongly violate CP and chiral symmetries
locally, producing in particularly $\pm 2N_f$ ( $\approx 6$) units
of chirality per sphaleron.
As pointed out in \cite{Dima},  such strong  local CP violation
induces special event-by-event fluctuations in the CP-odd
observables, e.g. they should induce  charge asymmetry along the
magnetic field.
 Clearly those should be looked at in  special studies.

 Another coherent color field configurations which deserve to be
specially studied are (colorelectric) flux tubes. In $pp$
collisions the view that a field created by longitudinally
separated charges makes a flux tube is a consequence of
confinement, and thus must happen in vacuum. Many popular event
generators are based on the Lund model, cleverly parameterizing
flux tube production and decay. In AA low energy collisions many
flux tubes are produced, and their possible fluctuations into  the
so called ``color ropes" has been studied, initiated by the paper
of \cite{ropes}. If two elementary color charges can be combined,
they may either cancel each other or produce higher
representations of the SU(3) group, in which case the rope energy
(and entropy, after its decay) is proportional to its flux
$squared$.  Further applications of these ideas for strangeness
production in AA collisions can be found in \cite{ropes2}.

Studies of the flux tubes lay dormant till recent discovery of the
so called ``hard ridge"  \cite{comment} by STAR collaboration at
RHIC. One possible origin of it  \cite{Shuryak:2009cy} is
hydro-carried longitudinal flux tube, created at the hard
collision point. This explanation may work provided the flux tubes
survive as such till freezeout: as was argued in that paper this
indeed should happen at the periphery of the fireball, where
matter is not far from the deconfinement transition, forming a
kind of ``dual corona" of the QGP fireball similar to Solar corona
full of flux tubes.

\section{Which correlations should one  study?}

\subsection{Central collisions: two versus many-particle correlators }

 Suppose a given event has certain (2-dimensional) distribution over transverse momenta
 of the secondaries. This distribution can be  decomposed  into Taylor series of the momenta $p_x,p_y$ or into
 the angular harmonics
\be {dN \over dp_t^2d\phi}= f(p_t)\left[ 1+\sum_{n>0} \left(2a_n cos(n \phi)+2b_n sin(n \phi)\right) \right]  \nonumber \\
a_n = < cos(n \phi)> ,  b_n = < sin(n \phi)> \hspace{1cm}
 \ee
 (Note that instead of the square bracket one can also use the exponent of the sum, which will include trivial higher order effects and enforces
 positivity of the distribution: but we would assume all $v_n$ to be very small, for simplicity.)
 Instead of using the $a,b$ pair, one may also introduce the modula and the phases
writing it as $2v_n cos[n(\phi-\xi_n)]$ with positive $v_n$. In
order to simplify subsequent formulae, we however prefer to write
it using the complex exponent \be {dN \over dp_t^2d\phi}= f(p_t)
\left( \sum_{n} v_n e^{in\phi-in\xi_n } \right) \label{eqn_single}
\ee where the sum goes over all integer $n$, positive and
negative, with $v_0=1$ and $v_{-n}=v_n$.

The sum of the harmonics presumably describe some interesting shapes. For example, those can be the sound circles
from point perturbations, leading to two-maxima shapes mentioned in the Introduction.
Unfortunately, there are multiple perturbations  in any event, and  the
 fluctuations we discuss are so small that they can only be studied by finding statistically significant small correlations,
  using large ($\sim 10^9$) sample of
available events.

Discussing how to do it let us, for simplicity,  first discuss a
subset of exactly central  collisions, with zero (negligibly
small)  impact parameter and thus exact azimuthal symmetry of the
event ensemble. Individual events would have of course some
fluctuations pointing to some directions $\xi_n$ , but their
absolute directions are arbitrary, not defined relative to
anything physical. Whatever fluctuation happens, it is clear that
its copy rotated by any angle should also exist and have the same
probability.
%
Say, if the elementary perturbation is local (delta-function-like in the transverse plane): its angular position
  in the transverse plane is
 the only meaningful azimuthal orientation.  Let us  express it as
 \be \xi_n=\xi_p +\tilde \xi_n \ee
where the tilde indicates the  angle
$relative$ to the perturbation.
  Obviously $\xi_p$   is a random variable,  changing from
event to event, and thus should be averaged out. It turns out that
the way the correlations work out is quite different for (i) the
two-body and (ii) the many-body (three or more) correlation
functions. Indeed, in order to get the two-body correlation
function one has to take a the square of the single-body
distribution (\ref{eqn_single})
\be  \sum_{n_1,n_2}v_{n_1}v_{n_2}  exp [ in_1\phi_1+in_2\phi_2 \nonumber \\
- in_1\tilde \xi_{n_1} - in_2\tilde\xi_{n_2} -i(n_1+n_2)\xi_p ]
\nonumber \ee and average it over
 $\xi_p$ randomly distributed over the circle.  As a result only terms satisfying \be n_1+n_2=0 \ee
survive, which has three important consequences. First, a double
sum collapses into a single sum with the squared amplitude
$\epsilon_n^2$. Second, it becomes a function of the angular
$difference$ $\Delta \phi=\phi_1-\phi_2$, as expected. And, last
but not least, all the {\em phases $\xi_n$ disappear}.

This facts are of course  well known , usually expressed as harmonics of  the 2-body correlator
\be C_2(\Delta\phi) =  <{d^2 N \over d\phi_1d\phi_2\
}>|_{\xi_p} \ee
\be  c_{n\Delta} = {\int d(\Delta \phi) C_2(\Delta\phi)  cos(n \Delta \phi
) \over \int d(\Delta \phi) C_2}=<v_n^2> \ee
So, the two-body correlator  provides
  $squared$ amplitudes  of the original harmonics, $averaged$ over the events.
This is e.g. how Alver and
Roland \cite{Alver:2010gr} and others have obtained   their estimates for the  ``triangular" flow.

 However,  the situation is different for manybody (three or more) correlation functions.
 Indeed, if the single-body distribution  (\ref{eqn_single}) is cubed (or raised into higher power),  one finds  a $triple$ sum in which  random
 perturbation direction appears as $exp[i(n_1+n_2+n_3)\psi_p]$. Averaging over it leads now to the condition
 \be n_1+n_2+n_3=0 \label{eqn_cond} \ee
  Eliminating e.g. $n_3$ one  finds the double sum
\be \sum_{n_1,n_2} v_{n_1}v_{n_2}v_{n_1+n_2} exp\{i [ n_1(\phi_1-\phi_3)+  n_2(\phi_2-\phi_3) \nonumber \\
-n_1(\tilde\xi_{n_1}-\tilde\xi_{n_1+n_2})-n_2(\tilde\xi_{n_1}-\tilde\xi_{n_1+n_2} ) ]\} \nonumber
\ee
 in which the relative orientations of the different harmonics are still present.  Taking corresponding
 moments over the observed angle differences $\phi_1-\phi_3,\phi_2-\phi_3$ one can single out the corresponding terms,
 and thus measure the combination
\be <v_{n_1}  v_{n_2} v_{n_3}cos(n_1\xi_{n_1}  +n_2\xi_{n_2} +n_3\xi_{n_3})> \ee
 in which three integers are subject to
 the ``resonance" condition (\ref{eqn_cond}).
  Therefore, since there are many moments of various many-body correlators,  much more complete data analysis is possible, ultimately
allowing to observe also the $phases$ of the harmonics. The price to pay is that the smallness of all harmonics now appear in the third
(or higher) power, so it is more difficult to get its values out of the statistical noise.

Let us now briefly discuss how comparison to theory should be done, assuming such averages are experimentally measured.
There are two steps to be done. First, using linear response hydro one can approximate it as
\be
<v_{n_1}  v_{n_2} v_{n_3}cos(n_1\xi_{n_1}  +n_2\xi_{n_2} +n_3\xi_{n_3})>  \hspace{1.5cm} \\
= ({v_{n_1} \over \epsilon_{n_1}}) ({v_{n_2} \over \epsilon_{n_2}}) ({v_{n_3} \over \epsilon_{n_3}})  <\epsilon_{n_1}  \epsilon_{n_2} \epsilon_{n_3}cos(n_1\xi_{n_1}  +n_2\xi_{n_2} +n_3\xi_{n_3})> \nonumber
\ee
Second, one should change from the flow angles to deformation angles using (\ref{angle_shift}). Note that in each three terms $n_i$
in numerator and denominator cancel, leaving only $3\pi$ or simply the sign change
\be cos(n_1\xi_{n_1}  +n_2\xi_{n_2} +n_3\xi_{n_3}) = \\
-  cos(n_1\psi_{n_1}  +n_2\psi_{n_2} +n_3\psi_{n_3}) \nonumber \ee
The resulting correlation of the
amplitudes and orientations of the initial state fluctuations can be calculated from initial state model, as we have done above
for  the first three harmonics resonance $1+2=3$ as well as $3+2=5$.

   In fact it is not necessary to take angular moments of the correlation functions.
Instead of lengthy but obvious general expressions, let us give a
simple but instructive example. We are interested in telling the
difference between the two-prong events predicted by sound
propagation and three-symmetric case of purely triangular flow.

\begin{figure}[t]
\begin{center}
\includegraphics*[width=6.cm]{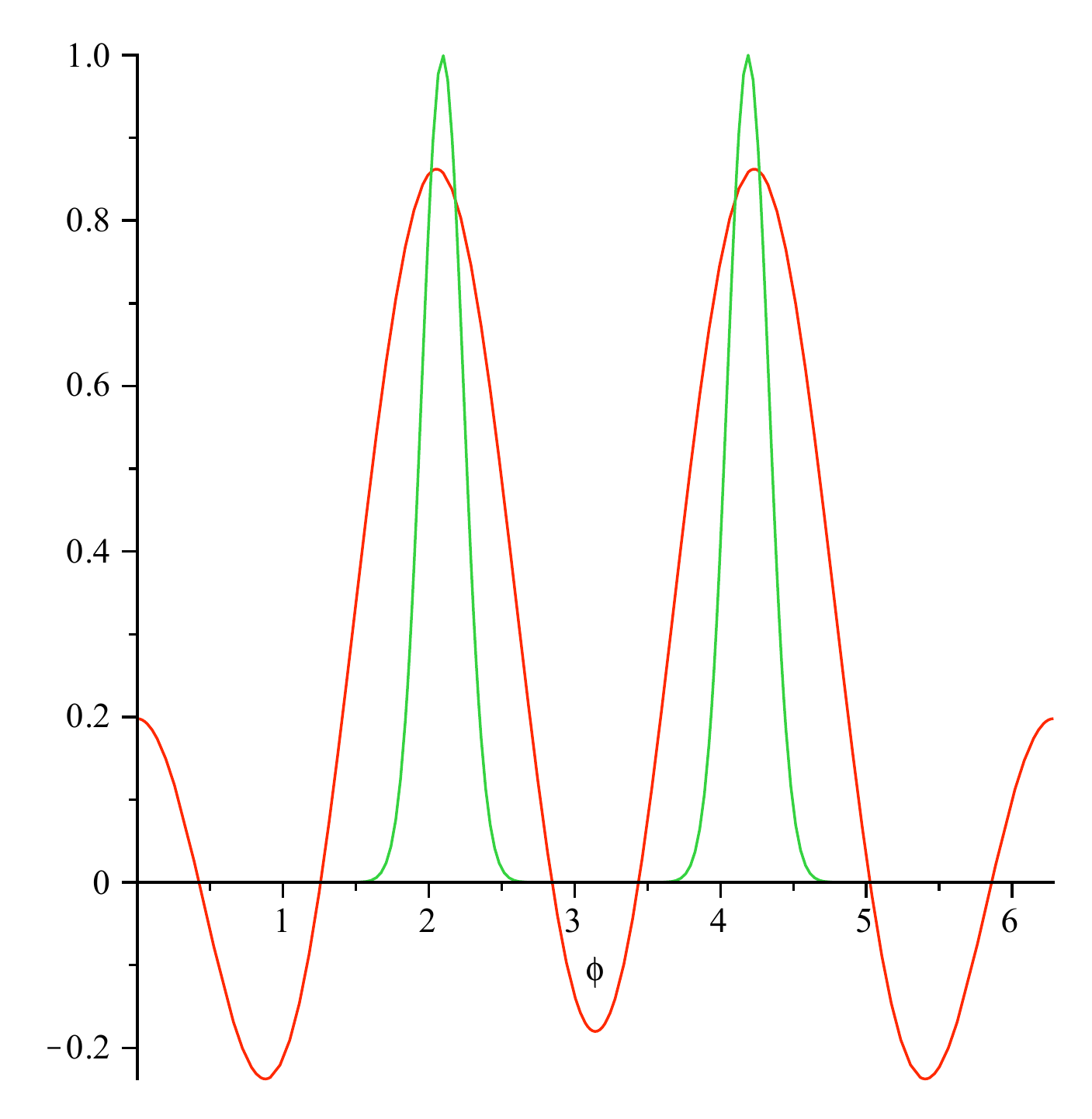}\\
\includegraphics*[width=6.cm]{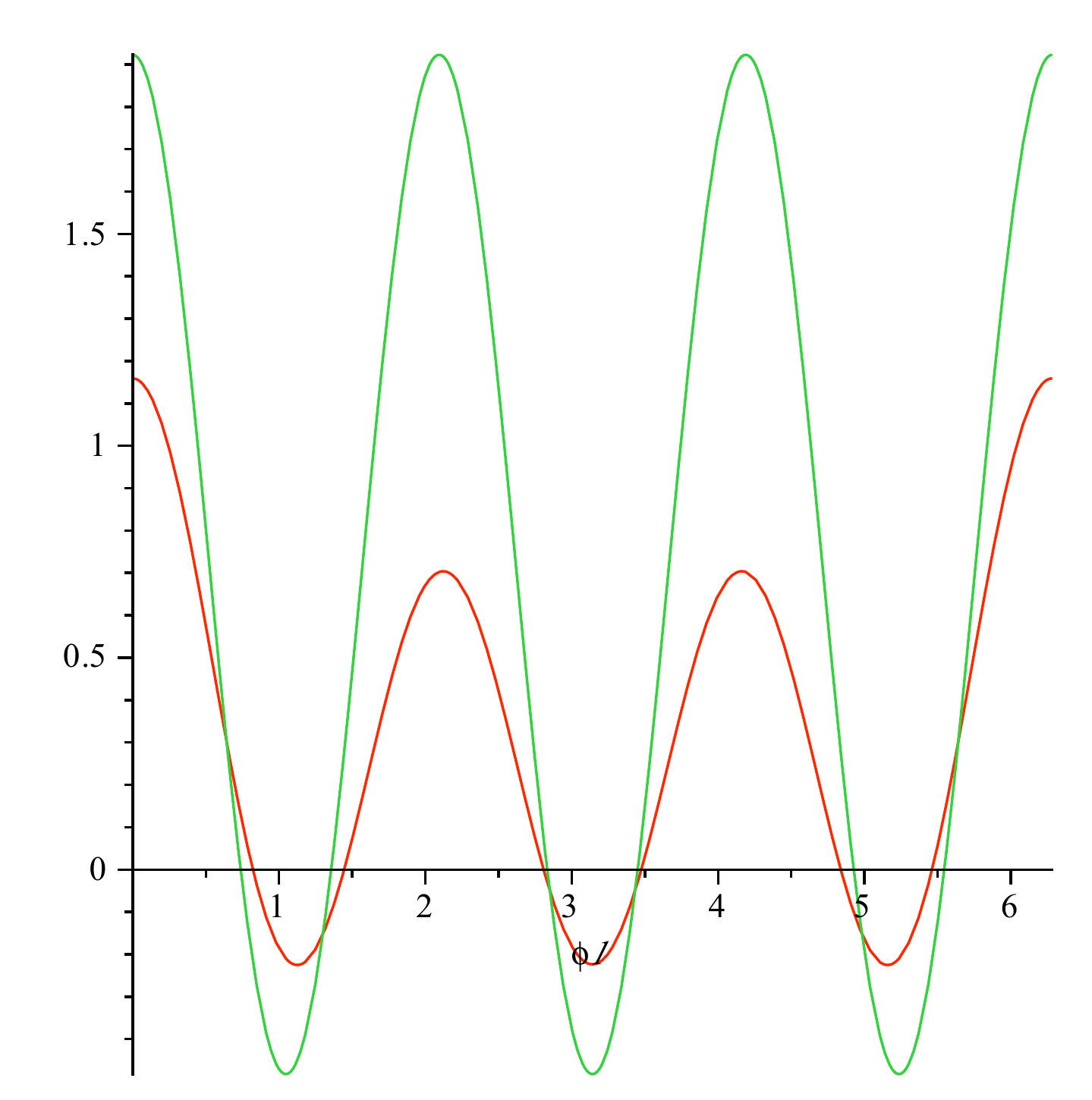}
\caption{ (upper) Two-peaked (green) curve shows the simplest
model of angular distribution to be discussed. The (red)
oscillating curve is its version after the filter removes all but
harmonics with $n>3$.\\
(lower) The lower (red) curve shows the dependence of the
2-particle correlation function on $\phi_{1}-\phi_{2} $ for the
filtered 2-peak distribution. The higher (green) curve is shown
for comparison, it corresponds to another model distribution with
$three$ identical peak shifted by 120$^o$. } \label{fig_2cor}
\end{center}
\end{figure}

    Consider the two simplest shapes depicted in Fig.\ref{fig_2cor}. The upper figure has (green) curve
which   shows the two-peak distribution, and the lower figure has
(red)  three-peak one. If it is projected into angular harmonics,
and the (viscous) filter kills all but the first three, one gets
their approximation by the other lines. The 1-st and 3-ed
harmonics are correlated because of absent third peak.

\begin{figure}[h]
\begin{center}
\includegraphics*[width=6.cm]{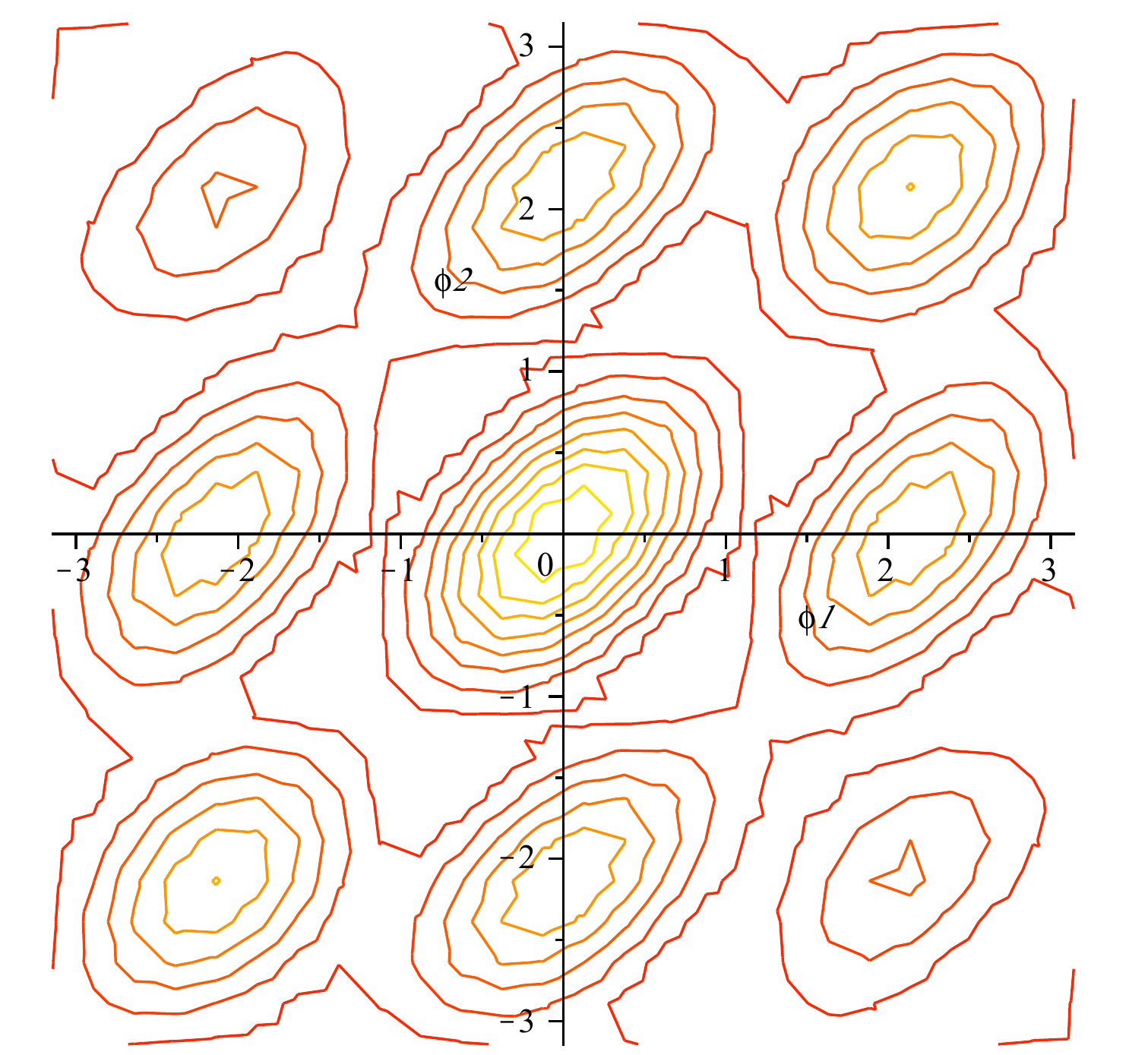}\\
\includegraphics*[width=6.cm]{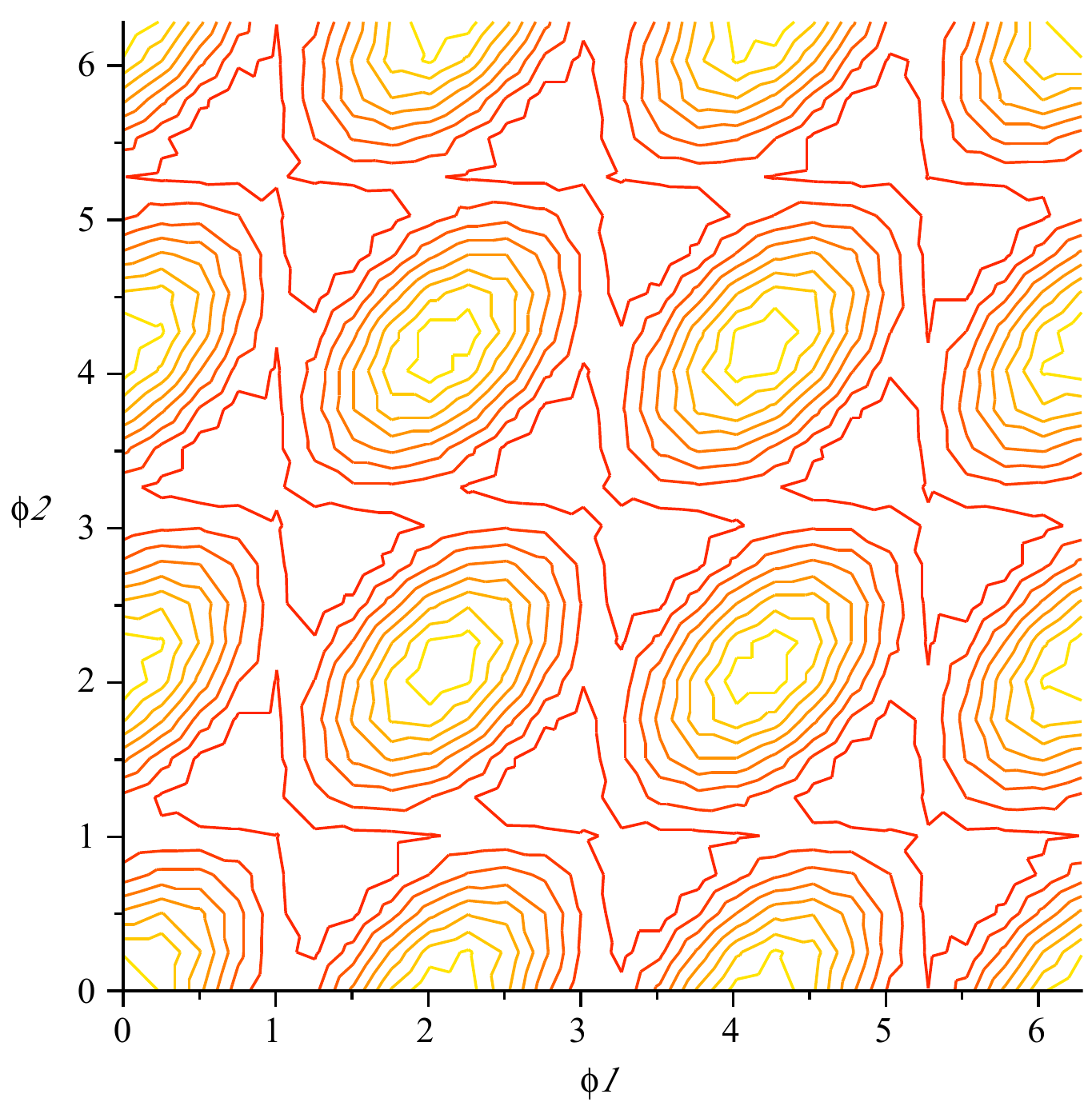}
\caption{The contour plot of the three particle correlators as a
function of $\phi_1-\phi_3$ and $\phi_2- \phi_3$, for 2-peak
distribution (upper figure) and 3-peak one (lower). As one can
see, their shapes are quite different. } \label{fig_3corr}
\end{center}
\end{figure}

 Defining
 the 3-particle correlator as for the $2$ particle one, with the averaging over the perturbation angle
 \be %
 C_{3}(\phi_{1},\phi_{2},\phi_{3}) =  <{d^2 N \over d\phi_{1}d\phi_{2} d\phi_{3} }>_{\psi_p}
 \ee %
 One finds it to be a function of two angles, say $\phi_{1}-\phi_{3}, \phi_{2}-\phi_{3}$. The corresponding plots for two examples in question, the 2-peak and 3-peak distributions, is shown
 in Fig.\ref{fig_3corr}. Comparing two plots one can see how different they are. In experiment, which will produce a mixture of the two,
 one would see that intermediate pattern, and hopefully determine the presence and probability of both type of events.

\subsection{Mid-central collisions and the two-body  correlators relative to the event plane}
Nonzero impact parameter violates  axial symmetry and creates
``directed flows", e,g, the famous elliptic flow with nonzero
$<v_2>\neq 0$. By mid-central collisions we mean a centrality
region in which $<\epsilon_2>$ is large itself, and is also large
compared to its fluctuations (recall that it is not so for central
and very peripheral collisions). For example, $\epsilon_2$ is 0.5
(0.3) for $N_p=100$ (200) participants, with $\delta
\epsilon_2\approx 0.1$. Furthermore, as seen in
Fig.\ref{angdist13}b, its angle $\psi_2$ is very much directed at
$\pm \pi/2$ (the tips of the almond) and therefore (using
(\ref{angle_shift} for $n=2$) the flow angle is peaked
``in-plane", $\xi_2=0,\pi$, as indeed observed.

If so, for one of the harmonics being the second e.g. $n_3=\pm 2$ one can approximate a product of three deformations as follows
\be
<v_{n_1}  v_{n_2} v_{n_3}cos(n_1\xi_{n_1}  +n_2\xi_{n_2} +n_3\xi_{n_3})>  \hspace{1.5cm} \\
\approx ({v_{n_1} \over \epsilon_{n_1}}) ({v_{n_2} \over \epsilon_{n_2}}) ({v_2 \over \epsilon_2})  <\epsilon_2>< \epsilon_{n_1}  \epsilon_{n_2} cos(n_1\xi_{n_1}  +n_2\xi_{n_2} )> \nonumber
\ee
by factoring out large and non-fluctuating $<\epsilon_2>$ from two other harmonics which are small and fluctuating.
Note that the resonance condition now means $n_1\pm 2=n_2$, and that
by putting $\xi_2=0$ we have selected a frame in which the (experimentally determined reaction plane) is the x axis.

    Basically the lesson here is that for mid-central collisions
the  ``reaction plane" plays the  same role as the third body, so
we are reduced to two small fluctuating and correlated harmonics.
 The simplest nontrivial example  of resonance condition of the kind is 3-1-2=0 
(recently studied by Teaney and Yan \cite{1010.1876}),
while the next is 5-3-2=0.

 We had already calculated the combinations of two fluctuating harmonics with
the appropriate cosines above, for these two cases, in the Glauber
model. They are by no means small: for example for the centrality class with 100
participants they are \be
 < \epsilon_1 \epsilon_3 \cos{(3\psi_3-\psi_1)}> \sim -0.015 \\
 < \epsilon_3 \epsilon_5 \cos{(3\psi_3-5\psi_5)}> \sim -0.05
 \ee
and therefore we expect it to be observable, with about as large
statistics as needed for the usual  quadratic fluctuations.

\section{Summary}


In this work we have (i) discussed the setting, identifying the
main scales of the problem. Then (ii) we studied in detail the
initial state fluctuations originating from nucleon positions,
emphasizing existence of the nontrivial phase relations between different harmonics.
Finally, (iii) we discussed correlations functions of 2 and many hadrons, pointed out the
principle difference between them, with the latter allowing to experimentally measure
the relative phases of these harmonics.

Let us now recapitulate the lessons from this study in a bit more detail.


Unfortunately, the perturbations we speak about are too small to
be measured directly, on event-by-event basis, and should be
instead reconstructed from the statistically obtained correlation
functions. One good thing coming from it is that multiple but independent
fluctuations from local perturbations in a single event and in the ensemble are
treated in one and the same correlation function, while all trivial effects are statistically subtracted.
Traditionally the initial state perturbations and final state corrections to collective flow
  are considered in a form of their angular harmonics, which we call $\epsilon_n$ and $v_n$ respectively.
  Their ratios $v_n/\epsilon_n$ can be calculated by the linearized hydrodynamics: the details of that is the subject of our next paper.

  ``Harmonic flows" have been so far treated as
independent or  incoherent fluctuations, added in quadratures. We
however pointed out that it is only true for say central
collisions and two-particle correlations. For central collisions
many (3 and more) particle correlations already include phases of
the deformations $\psi_n$, as do two-body correlations relative to
event plane for mid-central collisions. Many different moments
related to ``resonances" between 3 integers can be measured,
providing experimental opportunity to find out how the
perturbations are correlated and eventually to refine models of
the initial state. The magnitude of such terms, as we have shown
for Glauber model, is no smaller than for the terms already
studied.

 Coherence in phases of the deformations imply
the interferences between the harmonics of the flow. Only adding
them together one can follow how small ``hot" (or ``cold") spots
created by quantum fluctuations of interacting nucleons propagate
away from the point of origin. Only in this way one can understand
the role played by the ``hydrodynamical causality", insisting that
large part of the fireball  should remain completely unaffected by
the perturbation since the signal cannot possibly reach it before
the freezeout.
 Only a complete Green function, collecting all hydro harmonics. will describe correct
 shapes of propagating waves, such as e.g. (i)
the ``cylinders" from the initial state fluctuations, or (ii) the
``Mach cones" from quenched jets.

There are two basic scales defining those perturbations, the sound
horizon  $H_s$ (\ref{eqn_sound_horizon}) and the  ``the viscous
horizon scale" $R_v$ (\ref{eqn_visc_horizon}). The former gives
the size of the perturbation, stemming from a local perturbation,
the second its width.  We have for example argued that by changing
the centrality of the collisions, one can change the relation
between the (smaller) fireball size and the sound horizon: this
should dramatically change the shape of the event (see Fig.4 for
explanation).
It is an important objective of the experimental heavy ion program
in general to measure these two scales, extracting experimental
values of the speed of sound and viscosity. It can be done  for
example by changing centrality and observing the change of shape
of the underlying event.



\vskip .25cm {\bf Acknowledgments.} This work was
supported in parts by the US-DOE grant DE-FG-88ER40388.
Discussion with Frederique Grassi, explaining their finding in detail, have been very helpful. We also benefited from multiple discussion with
Derek Teaney.




\begin{thebibliography}{99}
\bibitem{comment} Note that ``hard ridge" should not be confused with the ``soft ridge". The latter, as discussed in this and other papers,
is naturally explained as a combined effect of ``harmonic flows" induced by Glauber fluctuations. Hard ridge, associated with the azimutal
direction of the hard trigger particle, does not yet have clear
\end{thebibliography}
\end{document}